\documentclass[twocolumn,prd,aps,floatfix,showpacs]{revtex4}
\usepackage{atlasphysics}
\usepackage{subfigure}
\usepackage{mathrsfs}
\usepackage{graphicx}
\usepackage{multirow}
\usepackage{rotating}
\usepackage{url}
\usepackage{amsmath}
\usepackage{hyperref}

\newcommand{\meff}{\ensuremath{m_{\mathrm{eff}}}} 
\newcommand{\mt}{\ensuremath{m_\mathrm{T}}}

\def\Ptmiss{\ensuremath{\vec{E}_\mathrm{T}^{\mathrm{miss}}}}

\newcommand{\oneleptitle}{Search for supersymmetry in final states with jets, 
missing transverse momentum and one isolated lepton in $\sqrt{s} = 7$ TeV $pp$ collisions using
1 fb$^{-1}$ of ATLAS data}

\newcommand{\hideit}[1]{{ }}

\newcommand{\btheta}{\boldsymbol\theta}


\begin{document}

\hspace*{13.8cm} \mbox{CERN-PH-EP-2011-153}

\hspace*{12.5cm} \mbox{Submitted to Physical Review D}

\vspace*{3mm}

\title{\oneleptitle}
\author{The ATLAS Collaboration}

\begin{abstract}
We present an update of a search for supersymmetry
in final states containing jets, missing transverse momentum, and one
isolated electron or muon, using $1.04$ fb$^{-1}$ of proton-proton collision data 
at $\sqrt{s} = 7 \TeV$ recorded by the ATLAS experiment at the LHC in the
first half of 2011.
The analysis is carried out in four distinct signal regions with either
three or four jets and variations on the (missing) transverse momentum
cuts, resulting in optimized limits for various supersymmetry models.
No excess above the standard model background expectation is
observed. Limits are set on the visible cross-section of new physics
within the kinematic requirements of the search. The results are
interpreted as limits on the parameters of the minimal supergravity
framework, limits on cross-sections of simplified models with specific 
squark and gluino decay modes, and limits on parameters of a model with
bilinear R-parity violation.
\end{abstract}

\maketitle

\section{Introduction and Analysis Overview}

Many extensions of the standard model predict the existence of new
colored particles, such as the squarks ($\squark$) and gluinos ($\gluino$)
of supersymmetric (SUSY) theories~\cite{susy}, which could be accessible at
the LHC. The dominant SUSY production channels are
assumed to be squark-(anti)squark, squark-gluino, 
and gluino-gluino pair production.
Squarks and gluinos are expected to decay to quarks and gluons
and the SUSY partners of the gauge bosons (charginos, $\, \chinopm$, and
neutralinos, $\, \nino$), leading to events with energetic jets.
In R-parity conserving SUSY models~\cite{r-parity}, the lightest supersymmetric
particle (LSP) is stable and escapes detection, giving rise
to events with significant missing transverse momentum.
In decay chains with charginos ($\squarkL \to q \chinopm$, $\gluino \to
q \bar{q}' \chinopm$), the chargino decay
can produce a high-momentum lepton.
LHC results of searches for SUSY with 35 pb$^{-1}$ of data collected in
2010, in final states with zero, one, or two leptons (where leptons
refer to either electrons or muons), can be found
in Refs.~\cite{atlas0lep,atlas1lep,atlas2lep1,atlas2lep2,atlasbjet,atlasgg,cms0lep,cmsgg,cms2lepOS,cms2lepSS,cmslg,cmsmultil,cmsbjet,cmsjets,cmssqgl,cms1lep}.
First results obtained with 1 fb$^{-1}$ of data collected in 2011 have
also been published for the no-lepton channel~\cite{cms2011}.

This paper reports on an update of a search for events
 with exactly one isolated
high-transverse momentum ($\pt$, relative to the beam direction) 
electron or muon, at least three 
high-$\pt$ jets, and significant missing transverse momentum ($\met$),
using data collected by ATLAS in the first half of 2011.
The analysis proceeds similarly to the analysis of 
the 2010 data~\cite{atlas1lep}, with a number of differences.
To cover a broader range of signals, the analysis has been
extended from one signal search region to four.
The kinematic requirements on leptons and jets have been modified,
to accommodate changing trigger requirements, minimize the overlap
with searches in other final states, and optimize the sensitivity
of the search.

As in the 2010 analysis, a combined
fit to the observed number of events in signal and background control regions
is used to search for an excess of events in the signal regions. 
The control regions normalize the backgrounds from $W$ and $t \bar{t}$
production. To estimate these backgrounds in the signal
regions, an extrapolation of the individual background components from
the control to the signal regions is performed. This is done
using transfer factors obtained from Monte Carlo (MC) simulations
that represent the expected ratio of events in the signal and
control regions for the various background processes.

The selection cuts are optimized based on samples of simulated events.
The cut optimization was performed 
not only in the MSUGRA/CMSSM (minimal supergravity/constrained Minimal
Supersymmetric Standard Model) framework~\cite{msugra,cmssm}, but also
for simplified models characterizing specific SUSY production and decay
modes. The results are interpreted in these MSUGRA/CMSSM and simplified
model frameworks, as well as in a model with bilinear R-parity 
violation (bRPV)~\cite{Roy:1996bua}.


\section{Models}
\label{sec:models}

In the MSUGRA/CMSSM model, supersymmetry is characterized
by universal scalar and gaugino mass parameters $m_0$ and $m_{1/2}$ and
a universal trilinear coupling parameter $A_0$, all expressed at the
GUT scale, the ratio of the vacuum expectation values of the two Higgs
doublets, $\tan \beta$, and the sign of the Higgs mixing parameter $\mu$.
In this paper, results are interpreted in terms of $m_0$ and $m_{1/2}$
for fixed values of $A_0 = 0$, $\tan \beta = 10$, and $\mu > 0$.
The interpretation is given for $\tan \beta = 10$ rather than for
$\tan \beta = 3$ as in our previous publication~\cite{atlas1lep}, 
since $\tan \beta = 3$ is increasingly
disfavored by the results of direct Higgs boson searches.
The influence of a variation of $A_0$ on the results is very small,
whereas high values of $\tan \beta$ ($> 30$) mostly affect the behavior
of the third generation of squarks and sleptons, for which dedicated
analyses are developed.
ISAJET~\cite{isajet} is used to calculate the SUSY particle mass spectrum
at the electroweak scale.
For illustration purposes, the expected signal distributions of the MSUGRA/CMSSM
model point $m_0 = 500 \GeV$, $m_{1/2} = 330 \GeV$, which is close
to the expected sensitivity limit, are shown in the 
figures of this paper.

Simplified models~\cite{simplified1,simplified2} are characterized
by well defined SUSY particle production and decay modes, and a 
minimal particle content for the final state under study. This can
be achieved by assuming that all SUSY particles not of interest to
a specific model are very massive and decouple. In order to achieve
a final state with leptons, the simplified models considered here
contain a chargino decaying to the lightest neutralino (LSP) and
an on-shell or off-shell W boson: $\chinopm \to W^{(*)} \nino$. The
chargino arises from the decay of a squark or a gluino, via one
of the following two models considered:
\begin{itemize}
\item In the mass hierarchy corresponding to 
sequential squark-chargino-neutralino decay, hereafter
called the squark model, the decay chain
$\squark \to q' \chinopm \to q' W^{(*)} \nino$ is assumed to have a 100\%
branching fraction, and only first- and second-generation squark-squark 
and squark-antisquark
production is considered. This is achieved by setting all other
SUSY particle masses, including those of third generation squarks,
to multi-TeV values. This model is characterized by three free
parameters: $m_{\squark}$, $m_{\nino}$, and $x = (m_{\chinopm} - m_{\nino})/
(m_{\squark} - m_{\nino})$.
\item In the gluino-chargino-neutralino model, hereafter called
gluino model, the decay chain
$\gluino \to q \bar{q} \chinopm \to q \bar{q} W^{(*)} \nino$ is assumed 
to have a 100\% branching fraction, and only gluino-gluino
production is considered. This is achieved by setting all other
SUSY particle masses, including those of all squarks,
to multi-TeV values. This model is also characterized by three free
parameters: $m_{\gluino}$, $m_{\nino}$, and $x = (m_{\chinopm} - m_{\nino})/
(m_{\gluino} - m_{\nino})$.
\end{itemize}
The assumption of massive third generation squarks in the squark model
is motivated by the fact that the phenomenology of light third generation
squarks (production of top and/or bottom squarks) is covered by
a separate dedicated analysis~\cite{atlasbjet}. For each choice of the three
free parameters in the simplified models, the sparticle mass spectrum
at the weak scale, and the sparticle decays are fully specified.
Simplified models are used to identify the limits of the effectiveness of 
the search, characterize a possible excess in data, and derive
limits. Constraints on a wide variety of models can be deduced from
limits on simplified models~\cite{simplified2}.

The MSUGRA/CMSSM model and the simplified models assume R-parity
conservation. Additionally, results are interpreted in a model
that allows for bilinear R-parity breaking terms in the
superpotential~\cite{Roy:1996bua}. Such terms lead to non-vanishing
vacuum expectation values for the sneutrinos which in turn induce
a mixing between neutrinos and neutralinos, thus providing a 
phenomenologically viable alternative to the origin of neutrino 
mass and mixing~\cite{Hirsch:2000ef, Takayama:1999pc}. 
In the study presented here, the R-parity violating
couplings are embedded in an MSUGRA/CMSSM SUSY production model. 
For a chosen set of MSUGRA parameters, the bRPV parameters are 
unambiguously determined under the tree-level 
dominance scenario~\cite{Grossman:2003gq} by fitting them to the 
neutrino oscillations data as described 
in Ref.~\cite{Carvalho:2002bq}. The neutralino LSP
is unstable and decays within the detector through decay modes that predominantly include 
neutrinos~\cite{Porod:2000hv}. 
Such decays along with the presence of neutrinos in SUSY decay chains 
such as $\chinopm \to \ell \nu \nino$
lead to significant missing transverse momentum.
However, this model was not used to optimize the selection. 
Only the muon selection is considered in this analysis
since in the leptonic decays of the LSP, the electron channels 
are highly suppressed in favor of the $\mu$- and $\tau$-producing modes. 
Scenarios leading to
a long lifetime ($c\tau \gtrsim 15~{\rm mm}$) of the LSP are not considered here.

\section{The ATLAS Detector}
\label{sec:ATLAS}

ATLAS~\cite{Aad:2008zzm} is a particle physics detector with a forward-backward symmetric
cylindrical geometry and near $4\pi$ coverage in solid angle~\cite{atlascoordinates}.
The inner detector (ID) consists of a silicon pixel detector, a silicon microstrip 
detector (SCT), and a transition radiation tracker (TRT). The ID is
surrounded by a thin superconducting solenoid providing a 2 T magnetic field, 
and by high-granularity liquid-argon (LAr) sampling electromagnetic calorimeters.
Hadron calorimetry
is provided by an iron-scintillator tile calorimeter in the 
central rapidity range. The end-cap and forward regions are instrumented with LAr
calorimeters for both electromagnetic and hadronic measurements. 
The muon spectrometer (MS) is based
on three large superconducting toroids arranged with an
eight-fold azimuthal coil symmetry around the calorimeters,
and a system of three stations of chambers for the
trigger and chambers for precise measurements.

\section{Monte Carlo simulation}

MC simulations are used to develop the analysis, extrapolate
backgrounds from the control to the signal regions, 
and to assess sensitivity to specific SUSY signal models.
Samples of $W$ and $Z/\gamma^*$ production with accompanying jets are
simulated with ALPGEN~\cite{alpgen}, using the 
CTEQ6L1~\cite{cteq6l} parton density functions (PDFs).
Top quark pair production is simulated with MC@NLO~\cite{mcatnlo} and the
next-to-leading order (NLO) PDF set CTEQ6.6~\cite{cteq66}, which is used 
for all NLO MC.
Single top production is simulated
with MC@NLO. Fragmentation and hadronisation for the ALPGEN and MC@NLO
samples is performed with HERWIG~\cite{herwig}, using JIMMY~\cite{jimmy} for the
underlying event. Diboson production is simulated with HERWIG, using the
MRST2007LO*~\cite{mrst2007modlo} modified leading-order PDFs.
SUSY signal samples in the MSUGRA/CMSSM model and for the simplified models
 are generated with HERWIG++~\cite{herwigpp},
normalized using NLO cross-sections determined by PROSPINO~\cite{prospino}.
The bRPV sparticle spectrum is calculated with 
SPHENO~3.1~\cite{Porod:2003um,Porod:2011nf}, the event generation is carried 
out by PYTHIA6~\cite{pythia} and the NLO cross-sections are also provided by PROSPINO.
The MC samples are produced using an ATLAS parameter tune of PYTHIA
and HERWIG/JIMMY~\cite{mc10} and
a GEANT4~\cite{geant4} based detector simulation~\cite{atlassimulation}. 
Detailed comparisons of MC predicted lepton reconstruction and
identification efficiencies to the corresponding measurements from data
are used to determine scale factors. These scale factors obtained
from specifically selected event samples, such as $Z \to \ell \ell$,
are then used to correct the MC prediction of efficiencies and
acceptances for both signal and background events.
The MC samples are produced with a simulation of multiple interactions 
per LHC bunch crossing (pile-up).
Differing pile-up conditions as a
function of the instantaneous luminosity of the LHC machine are taken
into account by reweighting MC events according to the mean number of
interactions expected.

\section{Object reconstruction}

Collision events are selected by requiring a reconstructed primary vertex with 
at least five associated tracks, consistent with the beam spot position.

Electrons are reconstructed from clusters in the electromagnetic calorimeter
matched to a track in the inner detector~\cite{electronefficiency}.
Several requirements on the track and clusters are imposed to select 
true electrons.  The ``medium'' electron selection, used in 
this analysis to estimate the contribution from non-isolated and
misidentified electrons and to veto on dileptonic events,
is based on calorimeter shower shape, inner-detector track quality,
and track-to-calorimeter-cluster matching. Electrons in the final
selection are required to pass the ``tight'' electron definition, which
adds a requirement on the ratio $E/p$, where $E$ is the calorimeter cluster energy
and $p$ is the track momentum, and detection of transition radiation in the TRT.
Furthermore, the electron is required to be isolated: the $\pt$ sum of
tracks within a cone of $\Delta R < 0.2$ around the electron candidate
(excluding the electron candidate itself) is required to be less than
10\% of the electron $\pt$. All electrons are required to pass
kinematic cuts of $\pt > 20 \GeV$ and $|\eta| < 2.47$. In addition, 
electrons with a distance to the closest jet of $0.2 < \Delta R < 0.4$
are discarded, where $\Delta R = \sqrt{(\Delta \eta)^2 + (\Delta \phi)^2}$.
For ``tight'' electrons, the $\pt$ requirement is raised to $25 \GeV$.

Preselected muons are either the result of a combined track in the
muon spectrometer and in the inner detector, or a muon spectrometer
segment matching with an extrapolated inner detector track~\cite{atlas-winclusive}.
The matched inner detector track must have
$\geq 1$ hit in the pixel detector, $\geq 1$ hit in the inner layer
of the pixel detector if the pixel detector module at that location is
operational, $\geq 6$ hits in the SCT, and fewer than two missing hits
on the track in pixel and SCT detectors.
For $|\eta| < 1.9$, at least 6 TRT hits are required, and the
number of TRT hits that are classified as ``outliers''
must be less than 90\% of the total number of TRT hits on the track.
The latter cut is also applied if $|\eta| \geq 1.9$ and at least 6 TRT
hits are on the track.
TRT outliers appear in two forms in the track reconstruction, as a straw 
tube with a signal but not crossed by the nearby track, or as a set of 
TRT measurements in the prolongation of a
track which, however, failed to form a smooth trajectory together
with the pixel and SCT measurements. 
These quality cuts are put in place to suppress fake tracks and 
discriminate against muons from hadron decays.
Muons with a distance to the closest jet 
of $\Delta R < 0.4$ are discarded.
In order to reject muons resulting from cosmic rays, tight
cuts are applied on the proximity of the muon trajectories to the
primary vertex (PV): $|z_{\mu} - z_{\mathrm{PV}}| < 5$ mm 
and $d_0 < 2$ mm, where $z_{\mu}$ is the $z$ coordinate of the 
extrapolated muon track at the point of closest approach to the primary
vertex, $z_{\mathrm{PV}}$ is the $z$ coordinate of the primary vertex,
and $d_0$ is the magnitude of the impact parameter of the muon
in the transverse plane. These preselected muons,
similar to the electron case, are used to quantify the contribution
from non-isolated muons and to reject events with additional muons, 
and are required to have $\pt > 10 \GeV$, and $|\eta| < 2.4$.
For muons in the final selection, the $\pt$ requirement is raised to
$20 \GeV$, and the
muon is required to be isolated: the $\pt$ sum
of tracks within a cone of $\Delta R < 0.2$ around the muon candidate
(excluding the muon candidate itself) is required to be less than $1.8 \GeV$.

Jets are reconstructed using the anti-$k_t$ jet clustering 
algorithm~\cite{antikt} with a radius parameter of 0.4. 
The inputs to the jet algorithm are three-dimensional clusters formed
from energy deposits in the calorimeter.
The jets are calibrated using $\pt$- and $\eta$-dependent correction
factors based on MC simulation and validated by test beam
and collision data studies~\cite{atlascalibration}.
Preselected jets are required to have $\pt > 20 \GeV$
and $|\eta| < 2.8$. Events with jets not passing jet quality criteria
against noise and non-collision backgrounds~\cite{atlas-jet-clean} 
are rejected. Jets within
a distance $\Delta R < 0.2$ of a preselected electron are rejected,
since these jets are likely to be electrons also reconstructed as jets.
For jets in the signal regions, the $\pt$ requirement is tightened to
$25 \GeV$ and to remove jets that are not associated with the hard
scattering of interest, jets with associated tracks are required to
pass the selection that at least 75\% of the summed $\pt$ of all 
associated tracks must come from tracks associated to the selected
primary vertex.

The occurrence of a $b$-tagged jet in the final state is used 
to distinguish between $t \bar{t}$ and $W$ events.
The reconstruction of $b$-tagged jets proceeds as for other jets,
apart from the requirement that $|\eta| < 2.5$, and that a
$b$-tagging algorithm exploiting both impact parameter and secondary
vertex information~\cite{btag2011}
tags the jet.
This algorithm has a 60\% efficiency for tagging
$b$-jets in a Monte Carlo sample of $t \bar{t}$ events, with a mistag rate
for light quarks and gluons of less than~1\%.

The missing transverse momentum $\met$ in this analysis is the opposite
of the vectorial $\pt$ sum of reconstructed objects in the event,
comprised of the jets
with $\pt > 20 \GeV$, the selected lepton, any additional identified non-isolated
muons, and three-dimensional calorimeter clusters with $|\eta| < 4.5$ not belonging to any of the
aforementioned object types.

During a part of the data-taking period, an electronics failure in the LAr
barrel EM calorimeter created a dead region in the second and third
layers, corresponding to approximately $1.4 \times 0.2$ in $\Delta \eta \times
\Delta \phi$. Events with an electron in this region are vetoed,
leading to loss of signal efficiency of about 1\%. 
The energy measurement for jets in the data in the problematic
region is underestimated. 
A correction to the jet energy is made using the energy 
depositions in the cells neighbouring the dead region, and this is
also propagated to $\met$.
The correction to the
jet energy amounts to a few percent for jets just touching the dead
region and reaches 40 percent for jets in the center of the dead region.
The contribution of jets in the dead region to $\met$ can be estimated
and is denoted as $\met ({\mathrm{hole}})$. 
Projecting this quantity on the direction
of $\met$ gives the quantity $\Delta \met ({\mathrm{hole}}) =
\met ({\mathrm{hole}}) \cdot \cos \Delta \phi ({\mathrm{jet}},\Ptmiss)$.
Events with $\Delta \met ({\mathrm{hole}}) > 10 \GeV$ and
$\Delta \met ({\mathrm{hole}}) / \met > 0.1$ are rejected.
This requirement rejects less than 0.5\% of the events in the
signal regions, and up to 2\% of the events in the control regions.

In the event selection, a number of variables derived from the
reconstructed objects are used. The transverse mass $\mt$ formed by $\met$
and the $\pt$ of the lepton ($\ell$) is defined as 
\begin{displaymath}
\mt\ =\sqrt{2\cdot p_{\mathrm{T}}^\ell\cdot\met\cdot(1 - 
\cos(\Delta\phi(\vec{\ell},\Ptmiss)))}.
\end{displaymath}
The effective mass $\meff$ is
obtained from objects in the event as the scalar sum
\begin{displaymath}
\meff\ = p_{\mathrm{T}}^\ell + \sum_{i=1}^{3(4)} p_{\mathrm{T}}^{\mathrm{jet}_i} + \met , 
\end{displaymath}
where $p_{\mathrm{T}}^{\mathrm{jet}_i}$ are the transverse momenta of the 
three (four) leading jets.

\section{Trigger and data selection}

The data were collected between March and July 2011.
The trigger system selects events online by requiring an
electron or muon trigger to fire.
The electron trigger selects electrons that deposit an amount of energy
corresponding to $E_{\mathrm{T}} = E \sin \theta > 20 \GeV$
in the calorimeter. The muon trigger requirement
determines a logical OR between a trigger that requires a muon with
$\pt > 18 \GeV$ and a trigger that requires a muon of looser quality with $\pt > 40 \GeV$
in the barrel; the OR of these two triggers increases the trigger
acceptance in the barrel.
The trigger efficiency is measured in the data. 
To assure good data quality, only runs in which all
subdetectors perform well are used, resulting in a data set corresponding to an
integrated luminosity of $1.04$ fb$^{-1}$, with an estimated
uncertainty of 3.7\%~\cite{lumi2011}.

\section{Event selection}

The kinematic selections start by requiring the presence of exactly one lepton
(electron or muon) with $p_\mathrm{T}>25\,\GeV$ in case of an electron 
and $p_\mathrm{T}>20\,\GeV$ for muons. If another lepton is reconstructed 
with $p_\mathrm{T}>20\,\GeV$
(``medium'' electrons) or $p_\mathrm{T}>10\,\GeV$ (preselected muons), the event is rejected in
order to minimize overlap with other analyses aimed at final states
with higher lepton multiplicities.

\begin{figure*}
\center
\includegraphics[width=0.47\textwidth]{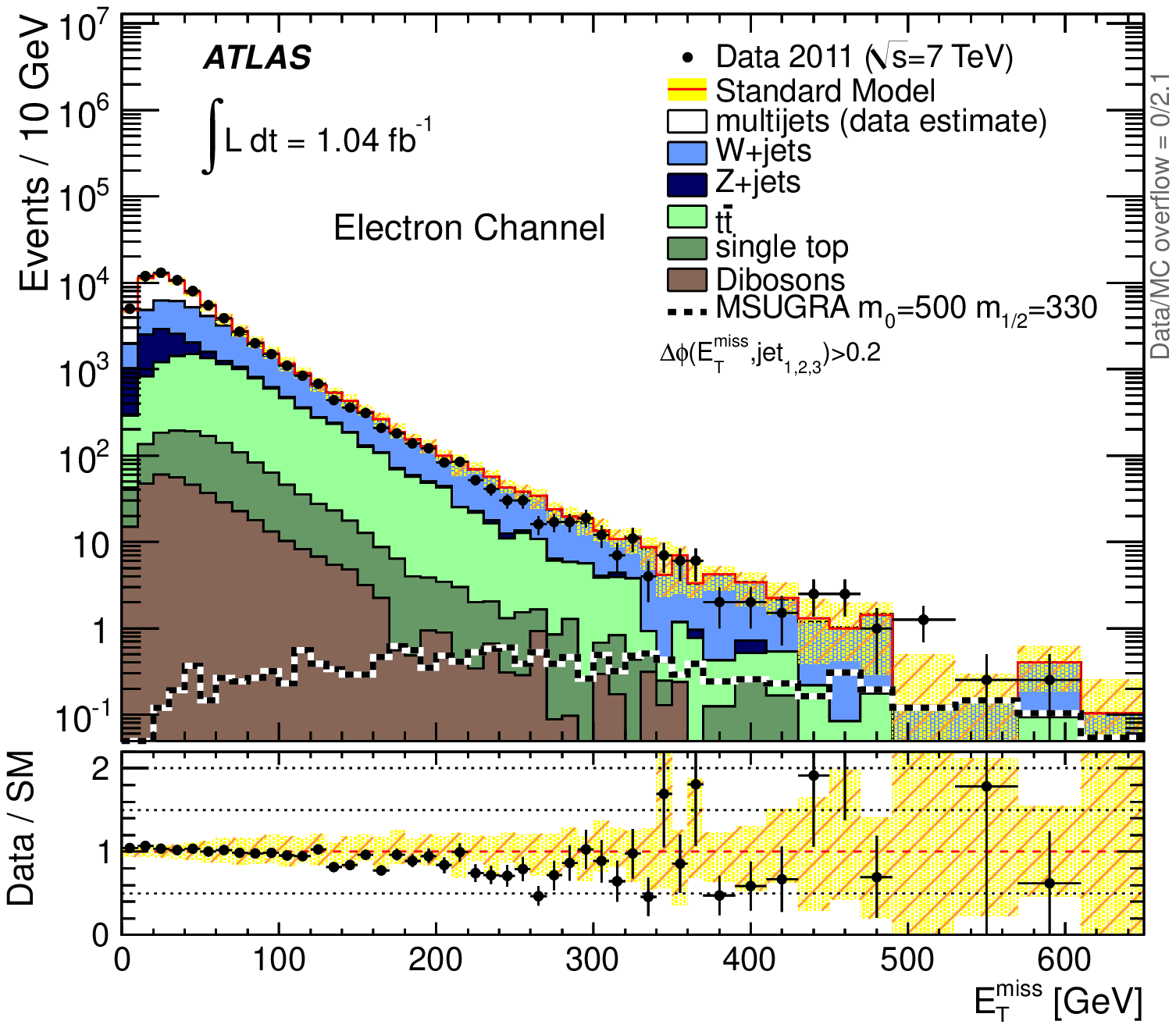}
\includegraphics[width=0.47\textwidth]{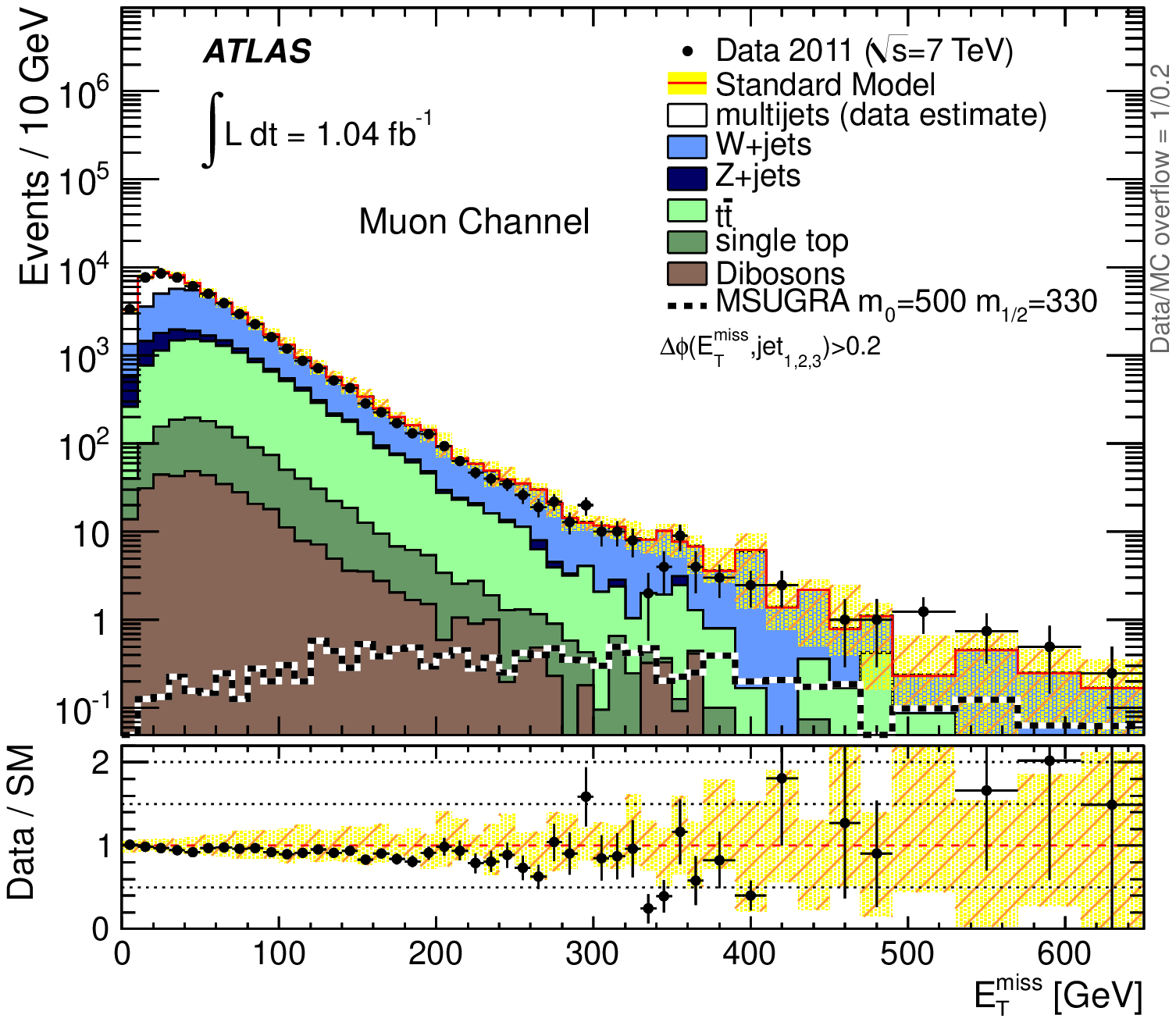}
\includegraphics[width=0.47\textwidth]{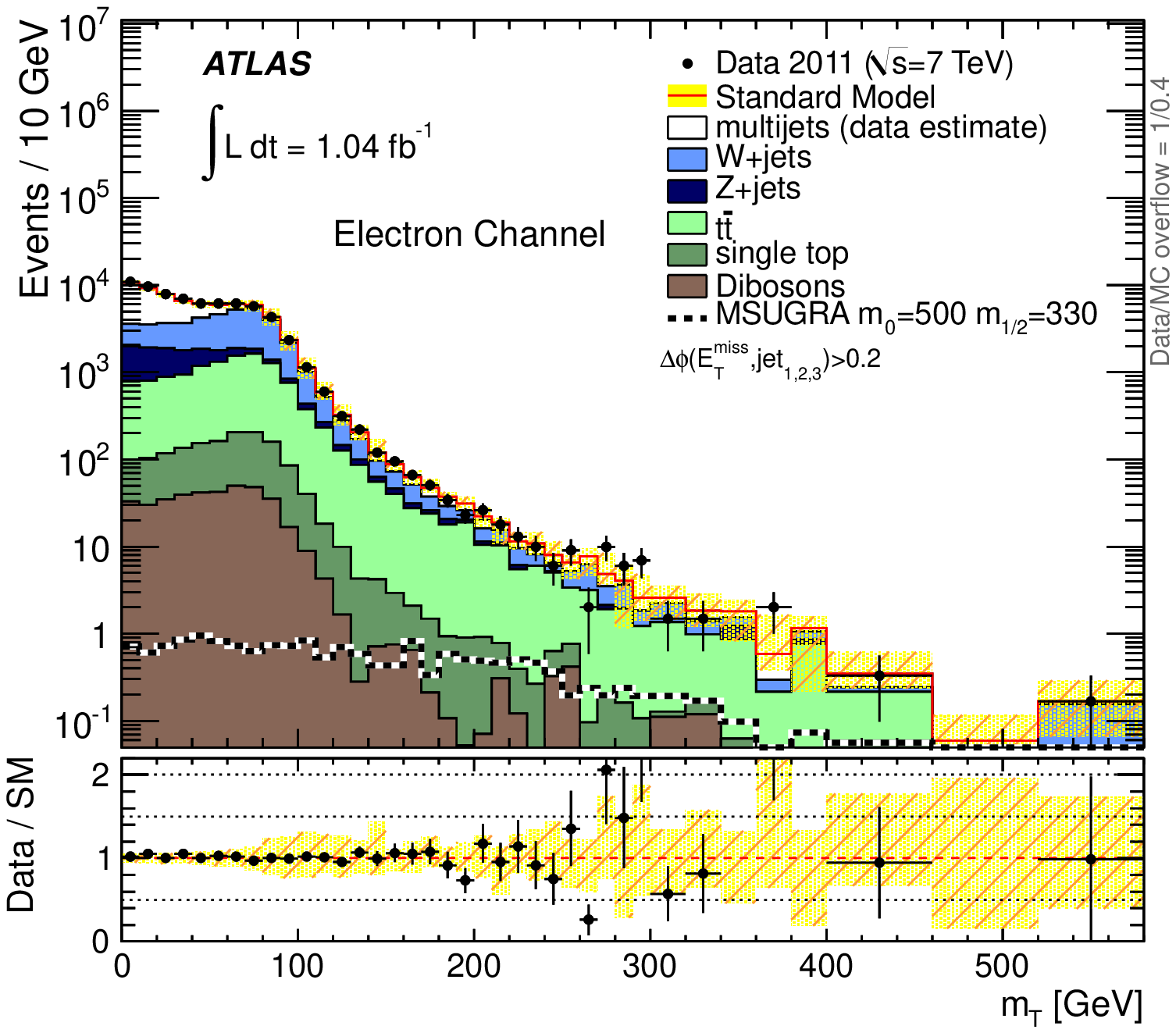}
\includegraphics[width=0.47\textwidth]{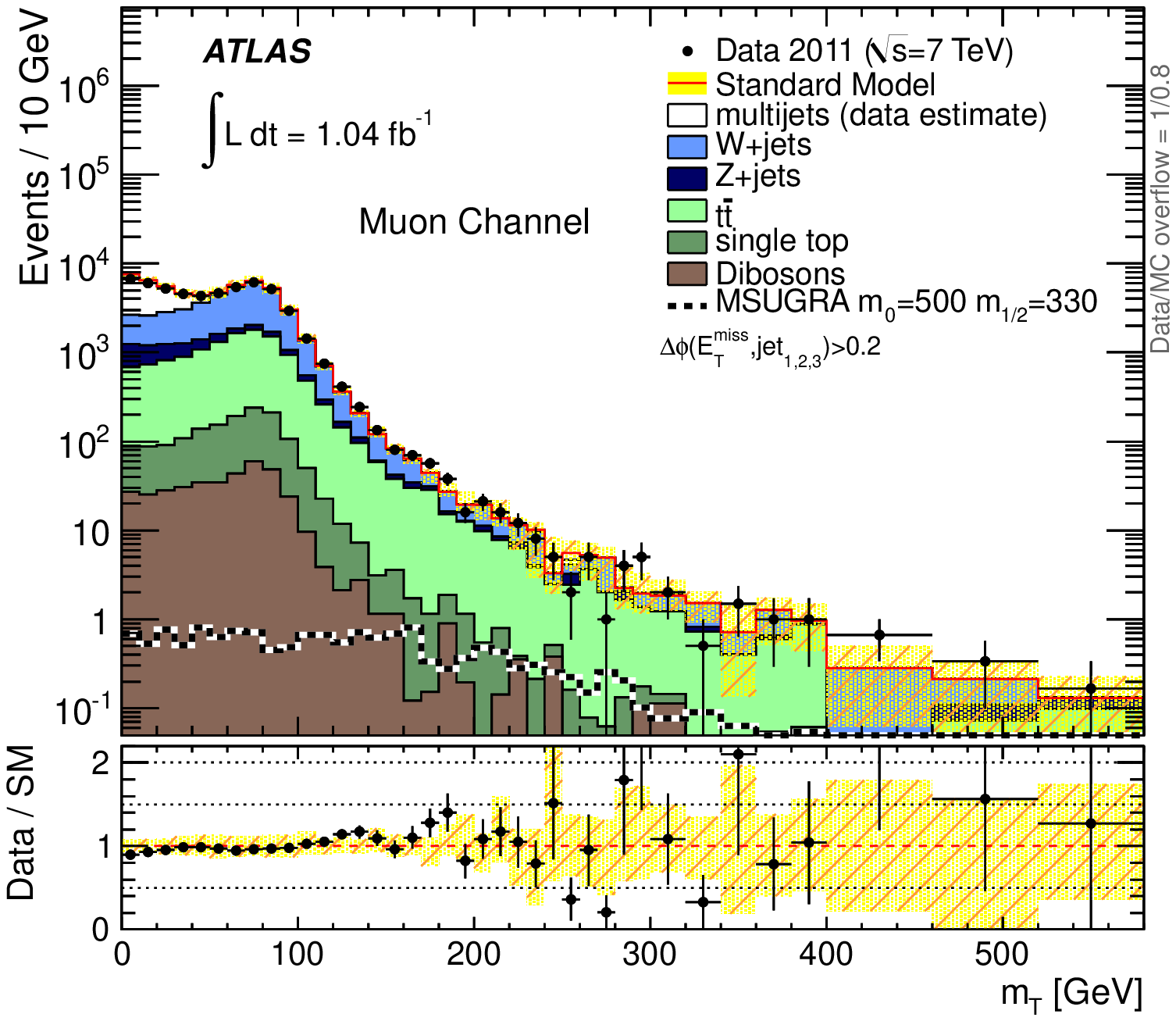}
\includegraphics[width=0.47\textwidth]{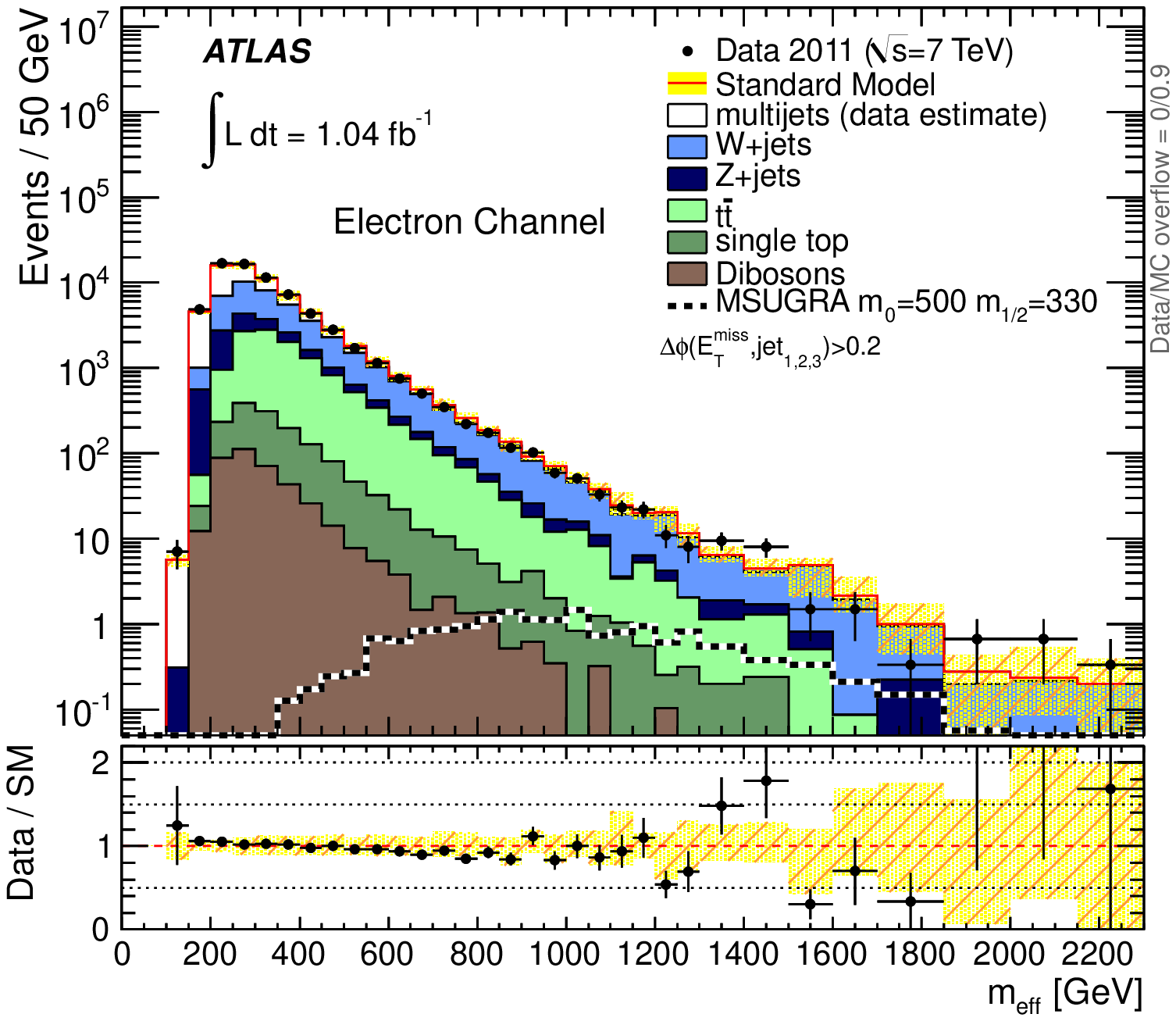}
\includegraphics[width=0.47\textwidth]{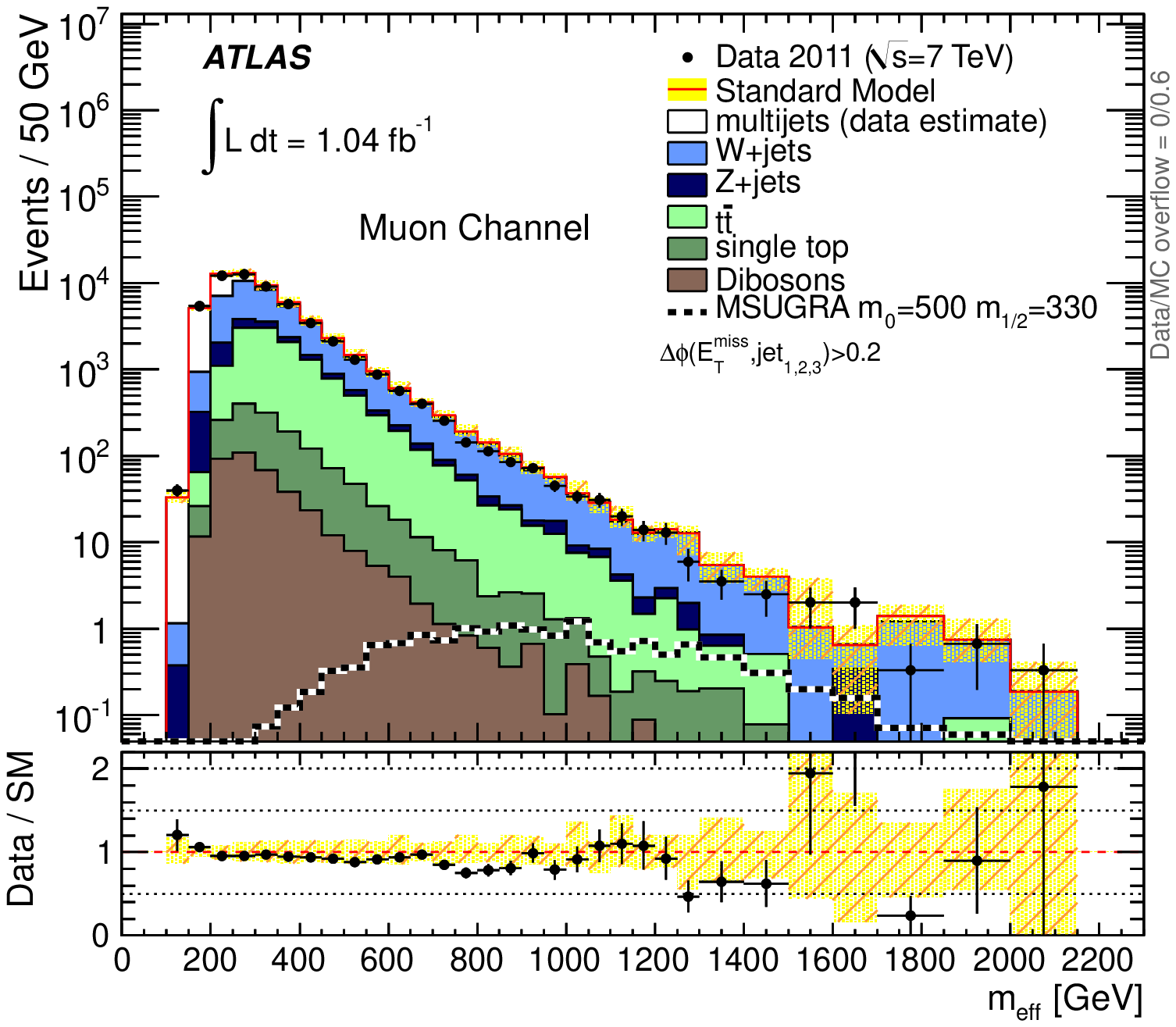}
\caption{
Distributions after requiring one electron with $\pt > 25 \GeV$ or
one muon with $\pt > 20 \GeV$, and at least three jets with
$\pt > 60, 25, 25 \GeV$ and $\Delta\phi({\mathrm{jet}}_{i},\Ptmiss) > 0.2$.
The top row shows the missing transverse momentum, $\met$, the middle row shows
the transverse mass, $\mt$, 
and the bottom row displays the effective mass, $\meff$. 
The electron channel is shown in the left column, the muon channel
is shown in the right column.
The ``Data/SM'' plots show the ratio between data and the summed 
standard model expectation.
In these plots, the standard model expectation is derived from Monte Carlo
simulations only, normalized to the theoretical cross sections.
The uncertainty band on the standard model expectation combines the MC statistical
uncertainty and systematic uncertainties on the jet energy scale and
resolution, the
lepton resolution and identification efficiencies, pile-up and luminosity.
For illustration, the expected signal distributions of the MSUGRA/CMSSM
model point $m_0 = 500 \GeV$, $m_{1/2} = 330 \GeV$ are also shown.
}
\label{fig:jetpresel3}
\end{figure*}

\begin{figure*}
\center
\includegraphics[width=0.47\textwidth]{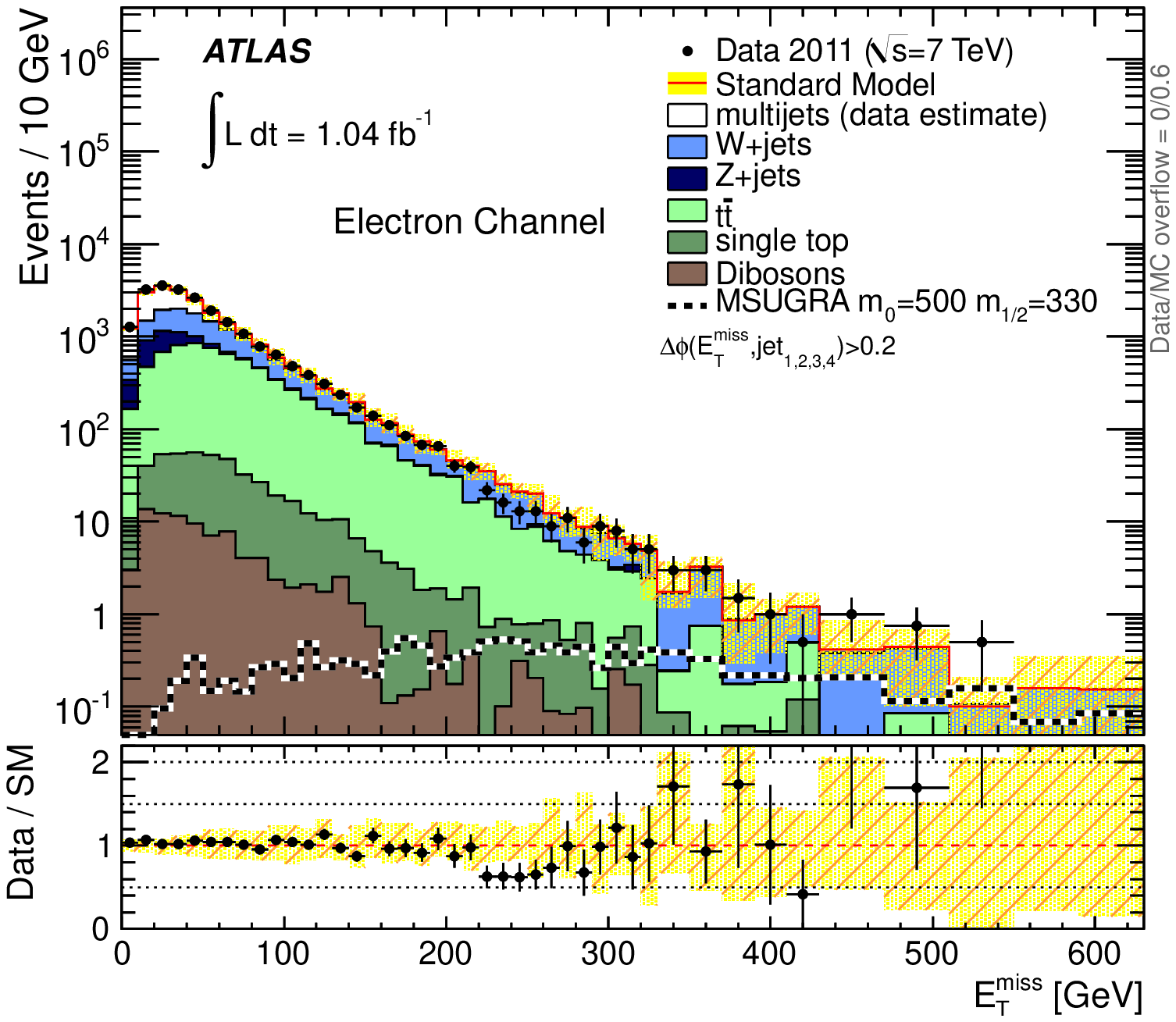}
\includegraphics[width=0.47\textwidth]{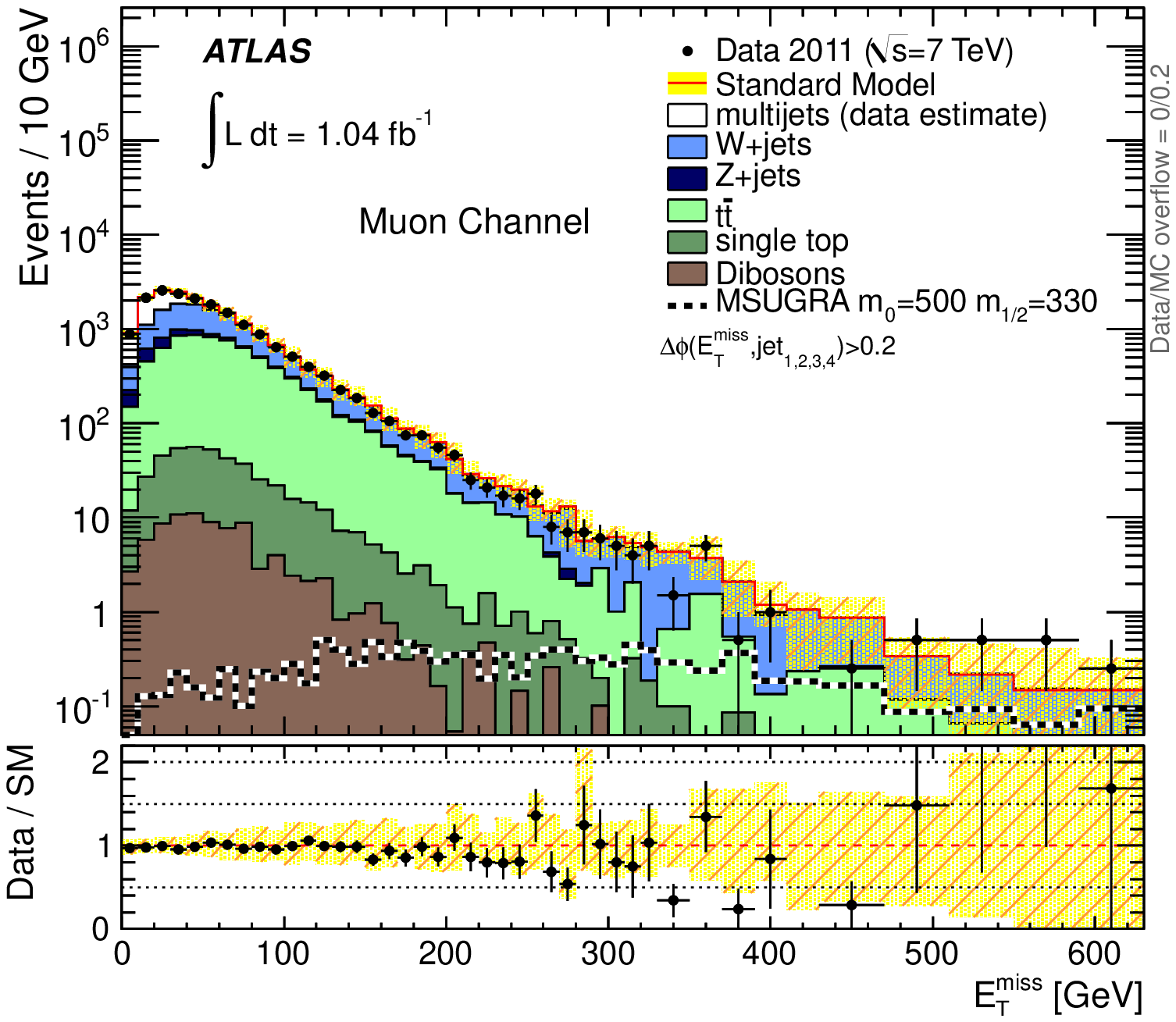}
\includegraphics[width=0.47\textwidth]{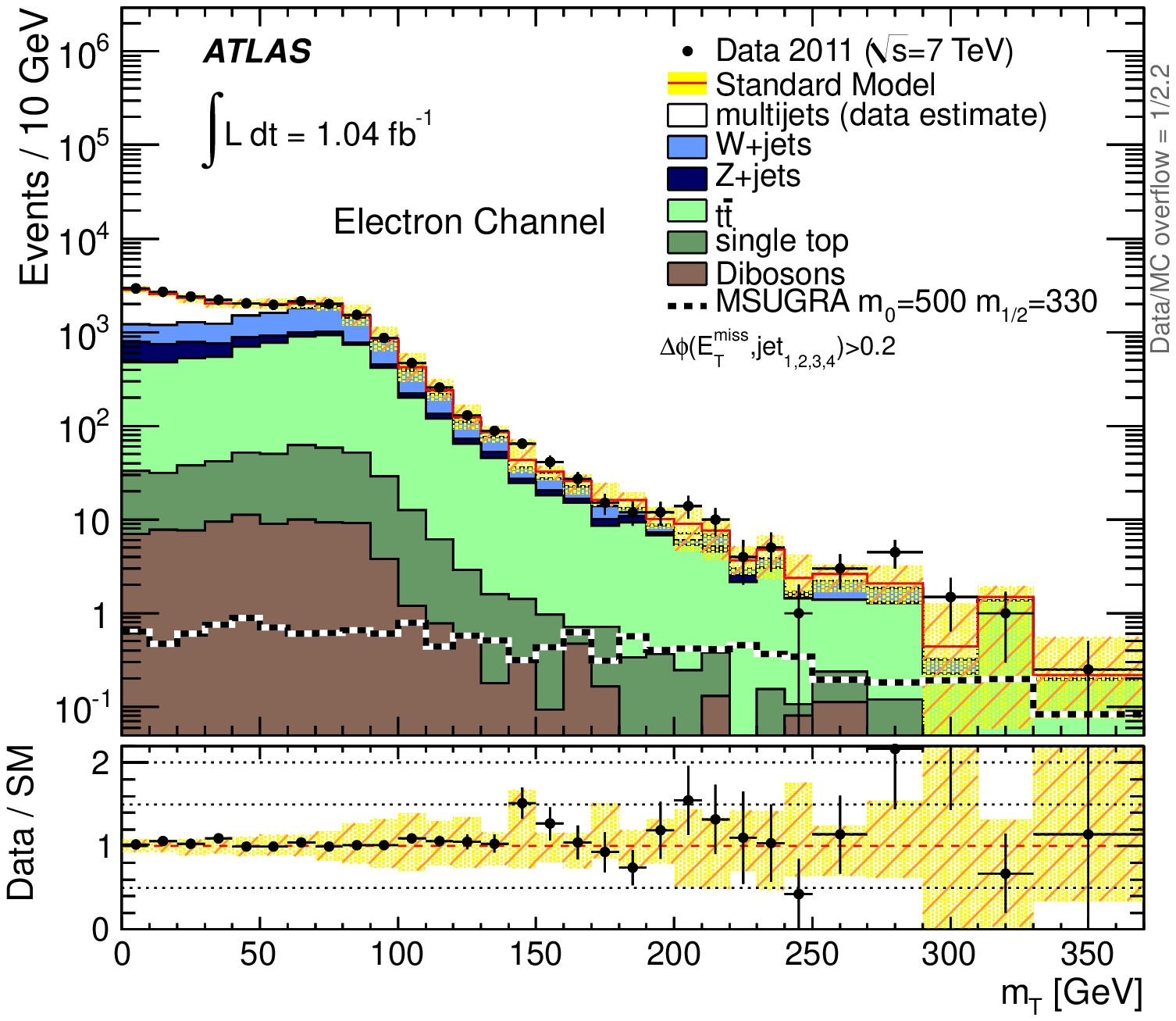}
\includegraphics[width=0.47\textwidth]{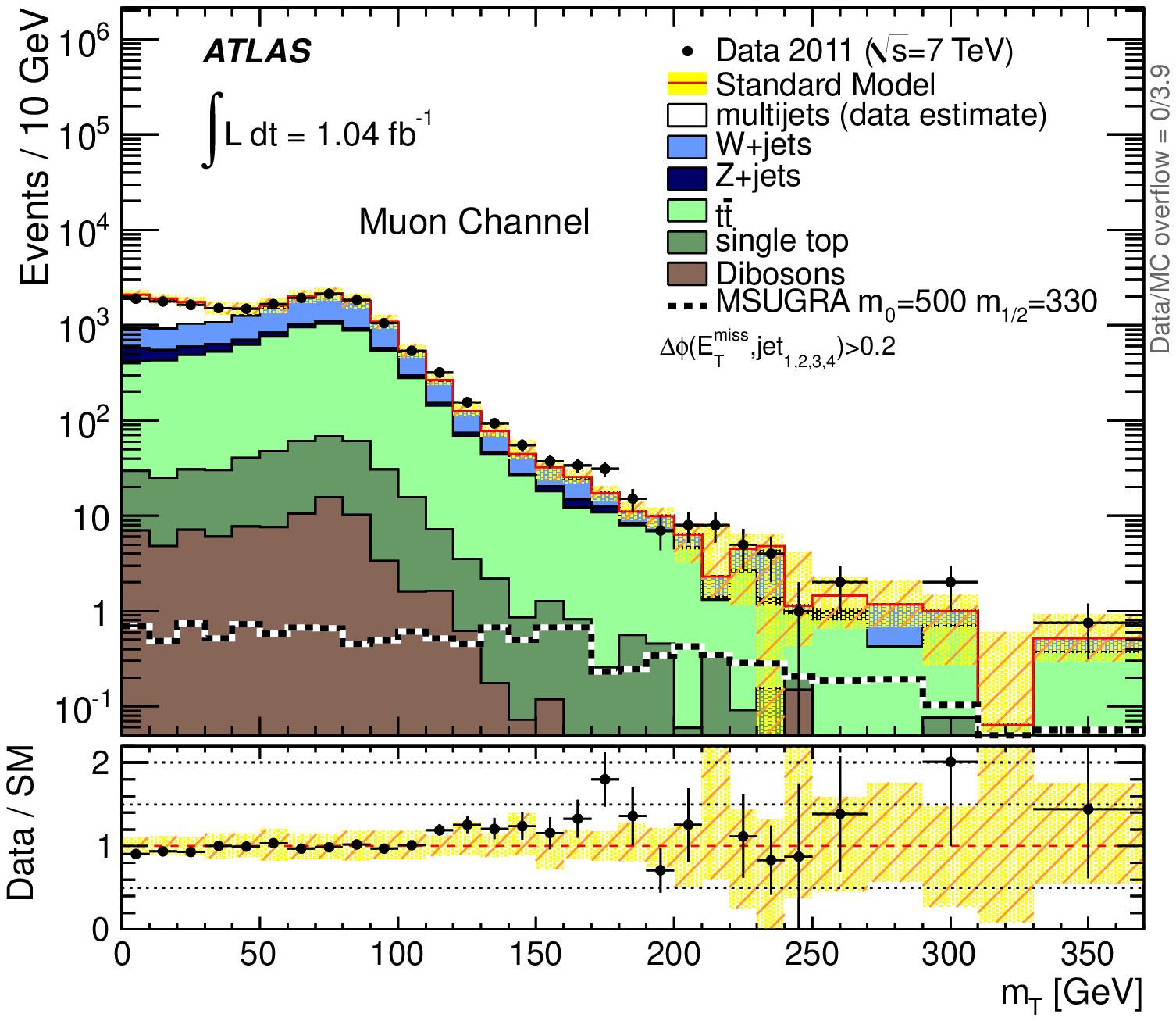}
\includegraphics[width=0.47\textwidth]{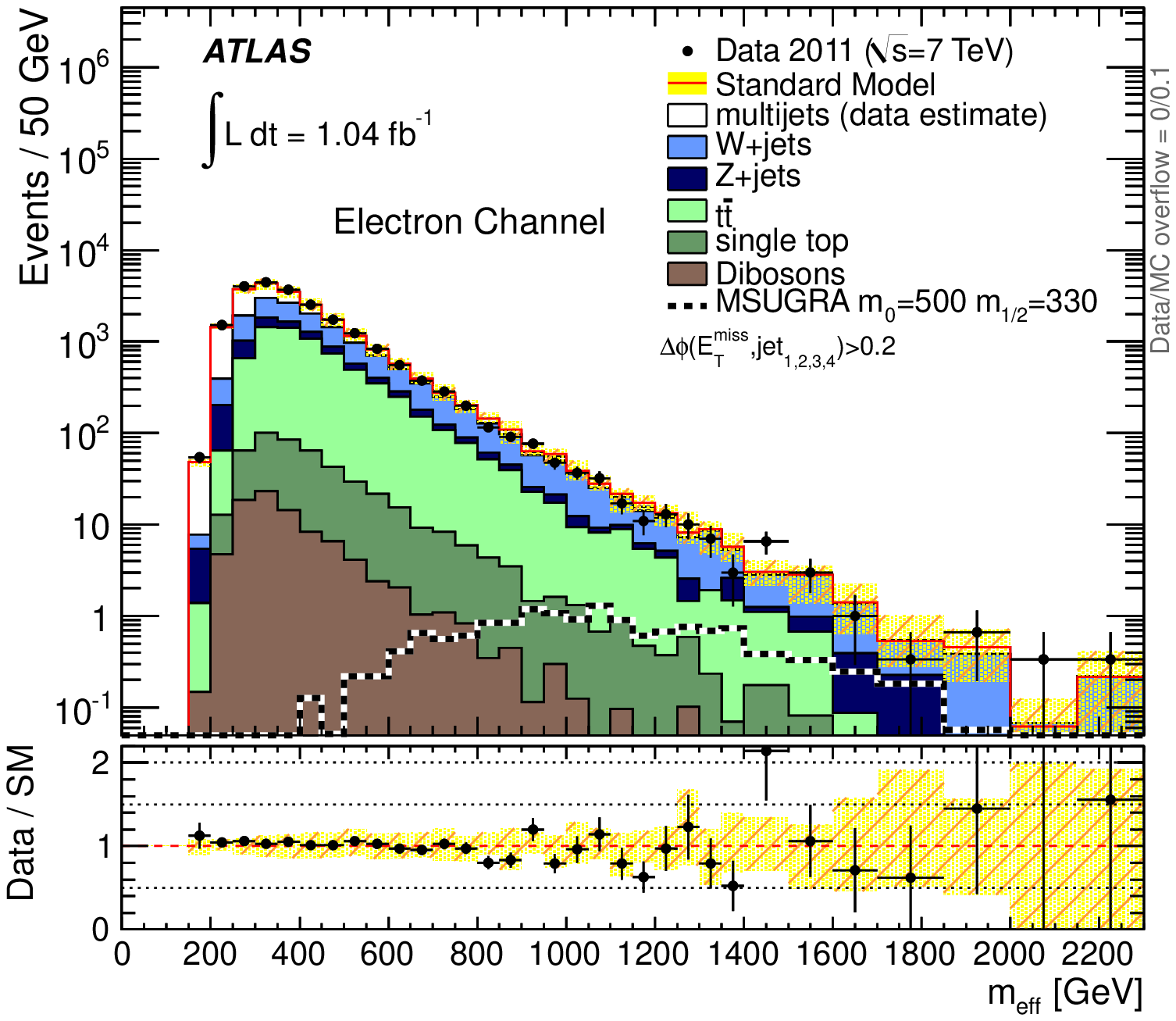}
\includegraphics[width=0.47\textwidth]{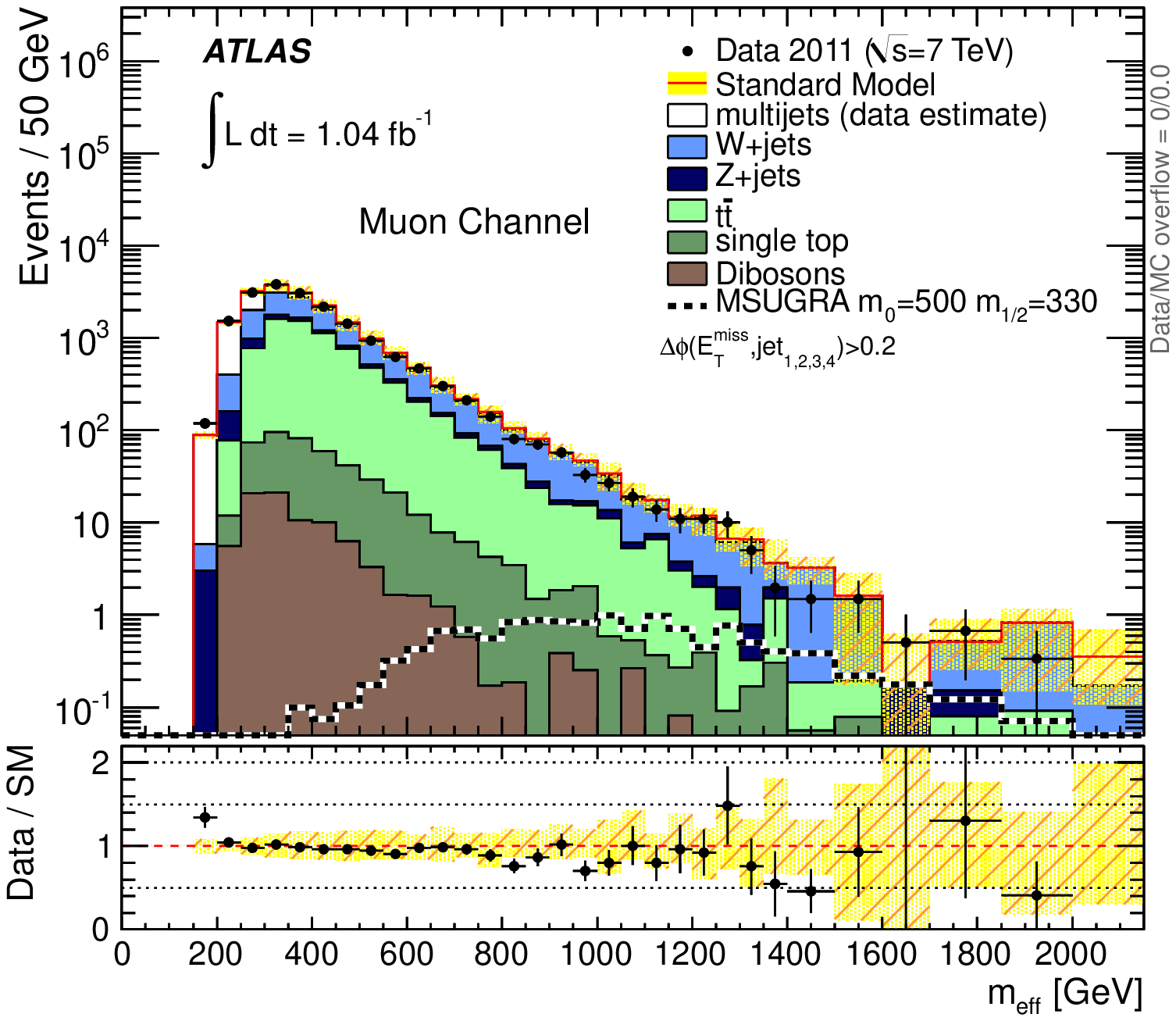}
\caption{
Distributions after requiring one electron with $\pt > 25 \GeV$ or
one muon with $\pt > 20 \GeV$, and at least four jets with
$\pt > 60, 25, 25, 25 \GeV$ and $\Delta\phi({\mathrm{jet}}_{i},\Ptmiss) > 0.2$.
The top row shows the missing transverse momentum, the middle row shows
the transverse mass, 
and the bottom row displays the effective mass. 
The electron channel is shown in the left column, the muon channel
is shown in the right column.
The ``Data/SM'' plots show the ratio between data and the summed 
standard model expectation.
In these plots, the standard model expectation is derived from Monte Carlo
simulations only, normalized to the theoretical cross sections.
The uncertainty band on the standard model expectation combines the MC statistical
uncertainty and systematic uncertainties on the jet energy scale and
resolution, the
lepton resolution and identification efficiencies, pile-up and luminosity.
For illustration, the expected signal distributions of the MSUGRA/CMSSM
model point $m_0 = 500 \GeV$, $m_{1/2} = 330 \GeV$ are also shown.
}
\label{fig:jetpresel4}
\end{figure*}

At least three or four good jets with pseudorapidity $|\eta|<2.8$ 
are required, depending on the selection, as outlined below.
Large mismeasurement of
the jet transverse momenta are avoided by requiring that \met\ is not aligned with
any of the three or four selected jets ($\Delta\phi({\mathrm{jet}}_{i},\Ptmiss) >
0.2$ ).
Kinematic distributions after application of the
lepton and jet selection requirements are shown in
Figure~\ref{fig:jetpresel3} for at least three jets and 
Figure~\ref{fig:jetpresel4} for at least four jets.

\subsection{Signal regions}
\label{sec:sigregdef}
Four different signal regions are defined to maximize the sensitivity to different 
kinematic configurations of supersymmetric event topologies.
\begin{enumerate}
\item {\it ``Loose'' 3-jet selection} (3JL).
The loose 3-jet selection is nearly identical to the selection used in the
analysis of the 2010 data~\cite{atlas1lep}. 
At least three jets, with
$\pt > 60 \GeV$ for the leading jet, and $\pt > 25 \GeV$ for the other
jets, are required. 
The transverse mass $\mt$ must exceed $100 \GeV$, and $\met$ must
be larger than $125 \GeV$. Two final cuts,
$\met/\meff > 0.25$ and $\meff > 500 \GeV$, define this signal region.
\item {\it ``Tight'' 3-jet selection} (3JT).
In the tight 3-jet selection, the requirement on the leading jet
$\pt$ is raised to $80 \GeV$. In addition to these cuts, the following 
criteria are applied: $\mt > 100 \GeV$,
$\met > 240 \GeV$, $\met/\meff > 0.15$ and $\meff > 600 \GeV$.
\item {\it ``Loose'' 4-jet selection} (4JL).
Four jets with 
$p_\mathrm{T}>25\,\GeV$ are required, with at least one of them exceeding $60\, \GeV$.
In addition to the jet cuts, the selection requires: $\mt > 100 \GeV$,
$\met > 140 \GeV$, $\met/\meff > 0.30$ and $\meff > 300 \GeV$.
\item {\it ``Tight'' 4-jet selection} (4JT).
A tight selection with at least four jets is defined. 
The $\pt$ requirement on the non-leading jets is raised to $40 \GeV$,
whereas the leading jet is still required to pass $\pt > 60 \GeV$.
To define this signal region, three more criteria are imposed:
$\met > 200 \GeV$, $\met/\meff > 0.15$ and $\meff > 500 \GeV$.
\end{enumerate}
The ``tight'' signal regions are optimized for the MSUGRA/CMSSM model,
which is characterized by energetic jets and large missing transverse
momentum. The ``loose'' signal regions perform better for the simplified
models with compressed particle spectra, i.e. when the LSP mass
approaches the squark or gluino mass. The 3-jet selection is optimized
for squark-squark and squark-antisquark production, the 4-jet selection
is better suited for squark-gluino and gluino-gluino production.

\subsection{Control regions}

Two classes of control regions (CR) are defined, 
i.e. separate control regions for the 3-jet
and the 4-jet selections.
The requirements on the lepton and the jets in the 
control regions are identical to those in the signal regions. 

\begin{figure*}
\center
\includegraphics[width=0.47\textwidth]{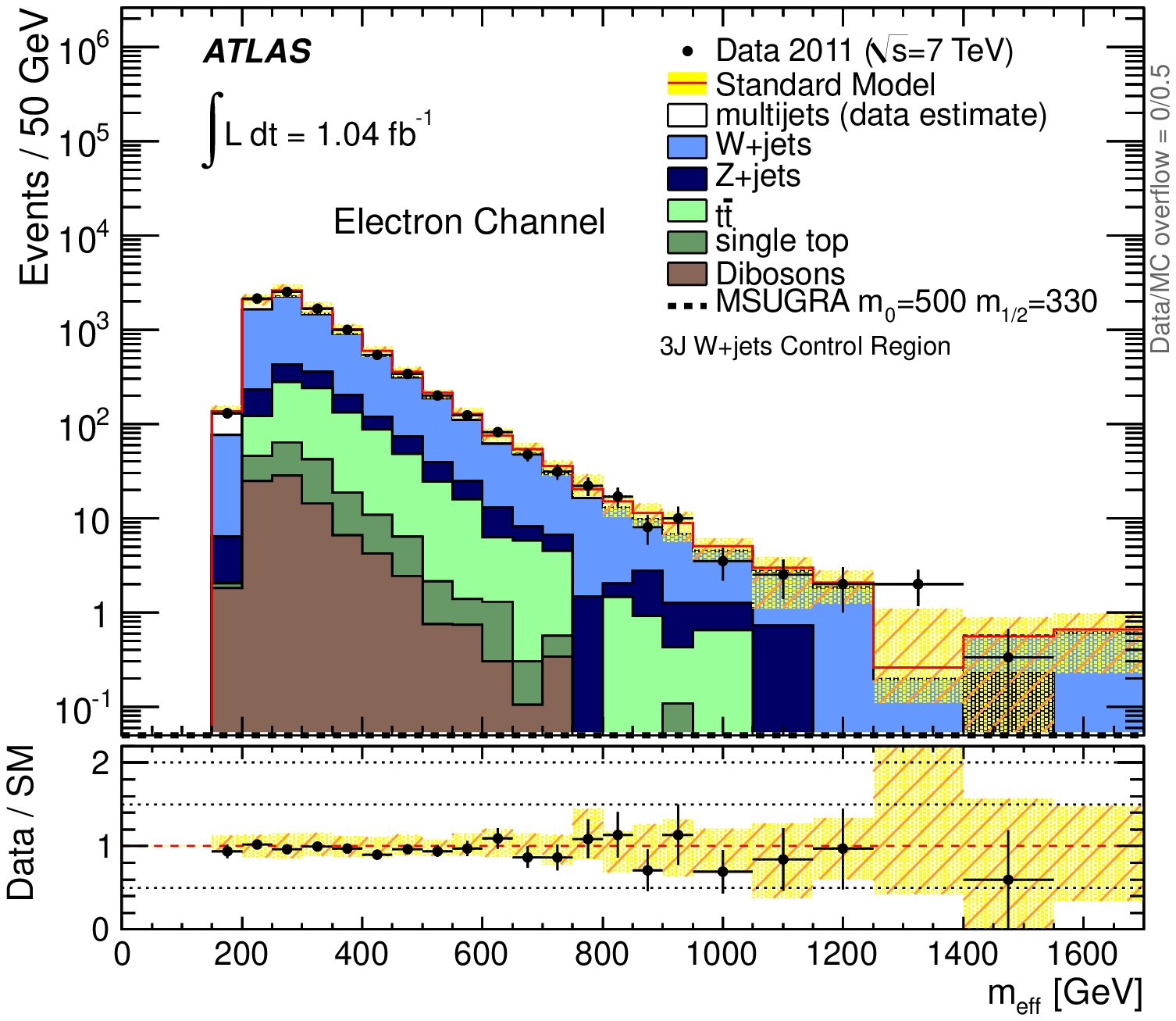}
\includegraphics[width=0.47\textwidth]{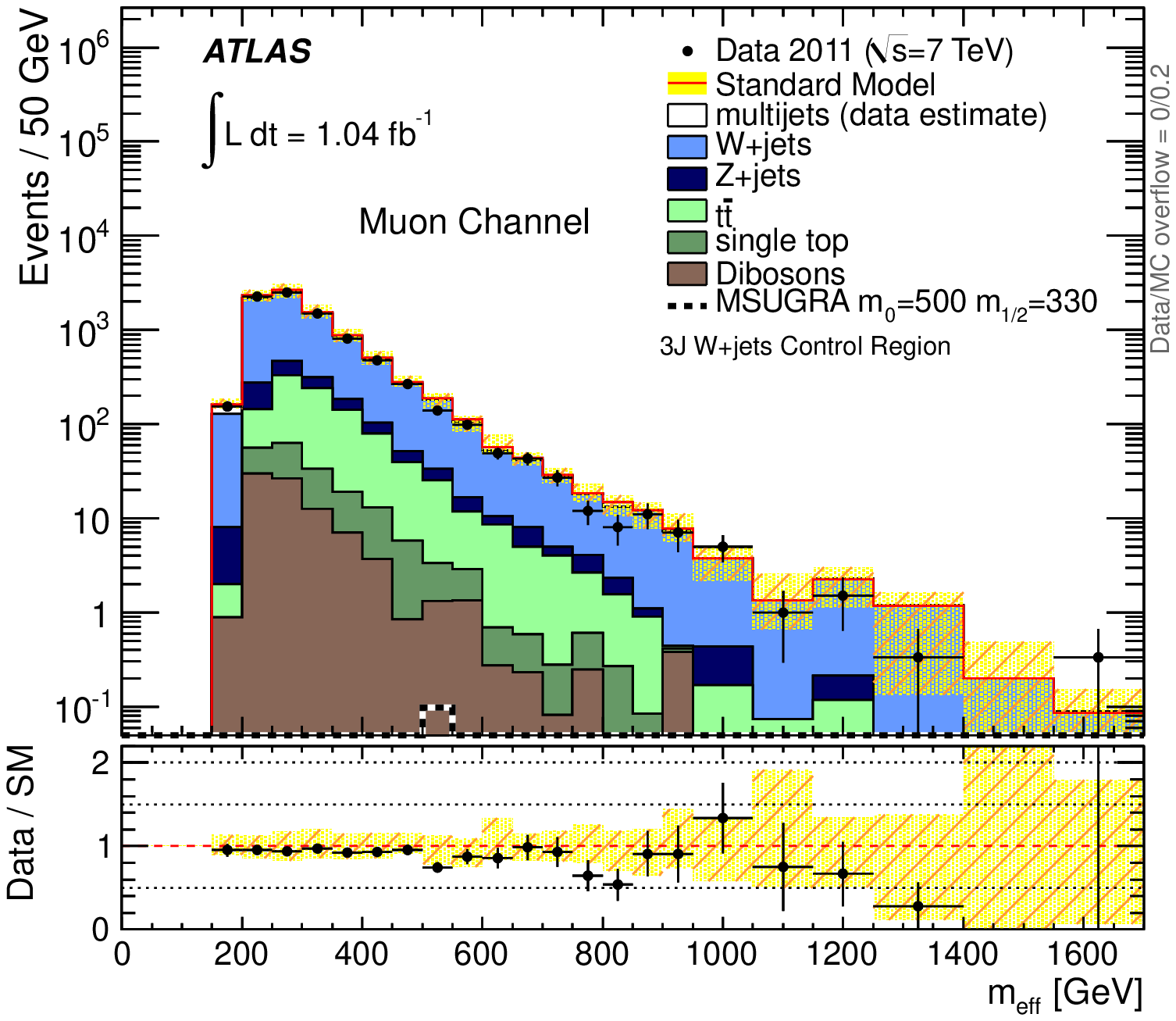}
\includegraphics[width=0.47\textwidth]{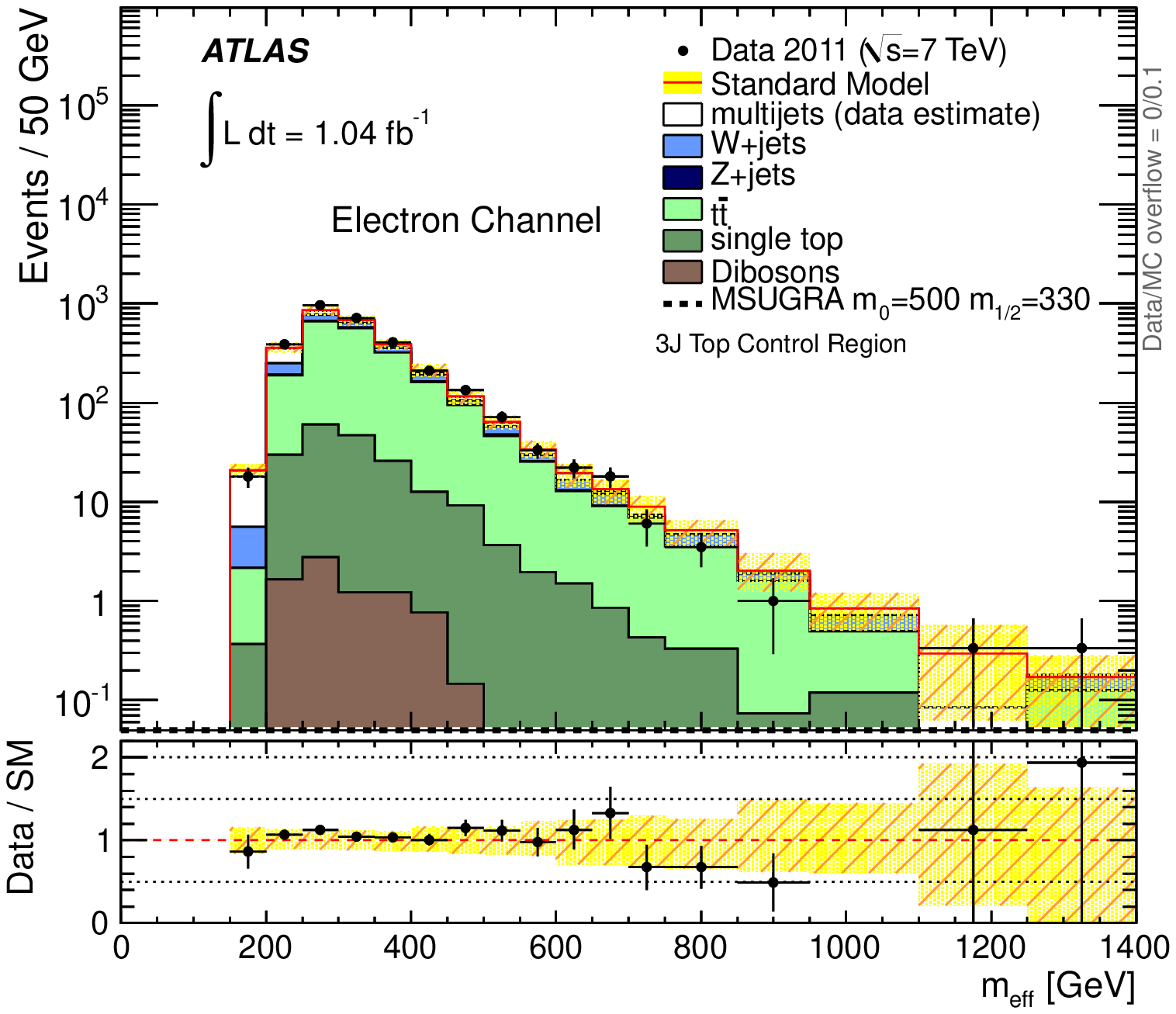}
\includegraphics[width=0.47\textwidth]{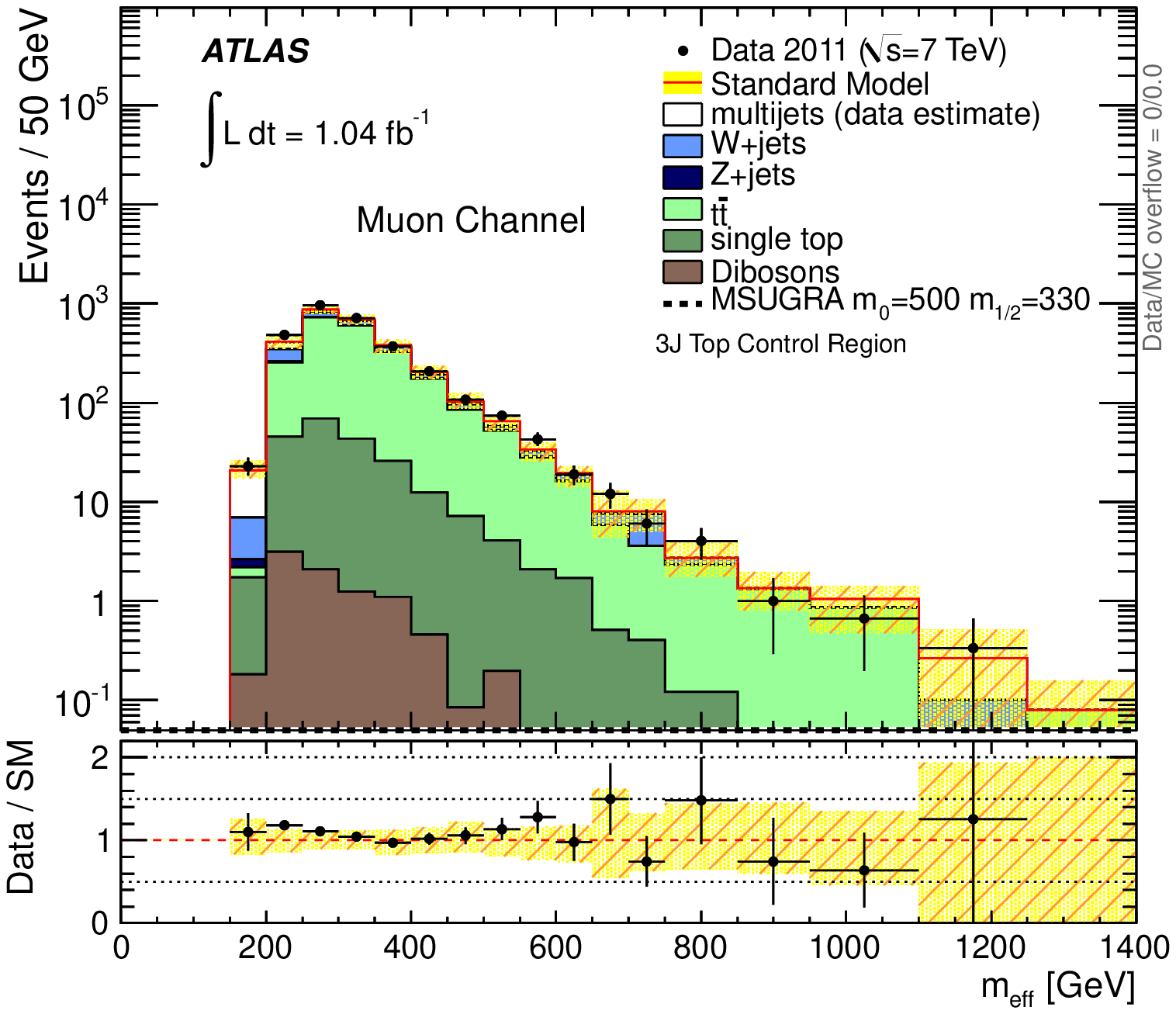}
\includegraphics[width=0.47\textwidth]{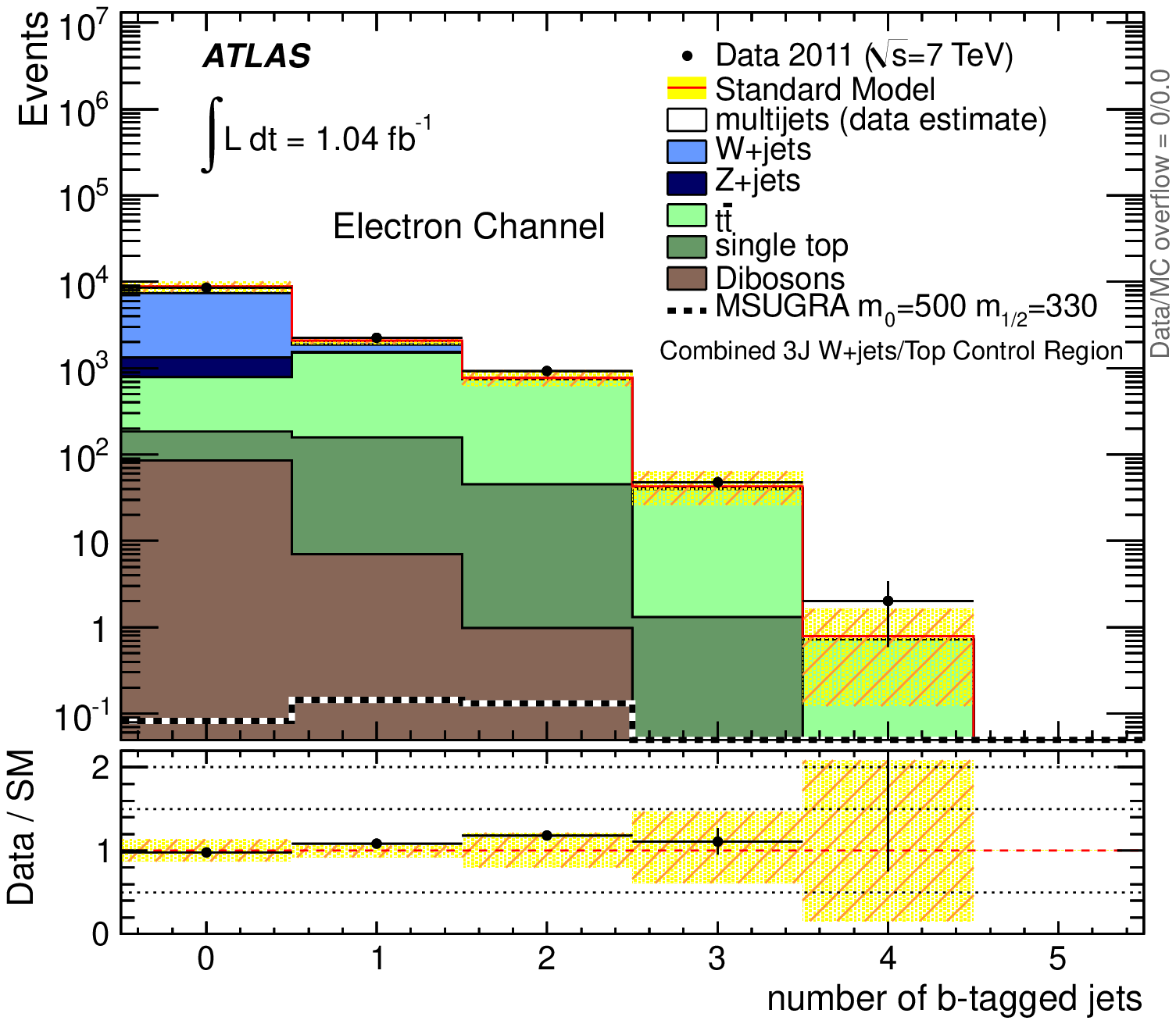}
\includegraphics[width=0.47\textwidth]{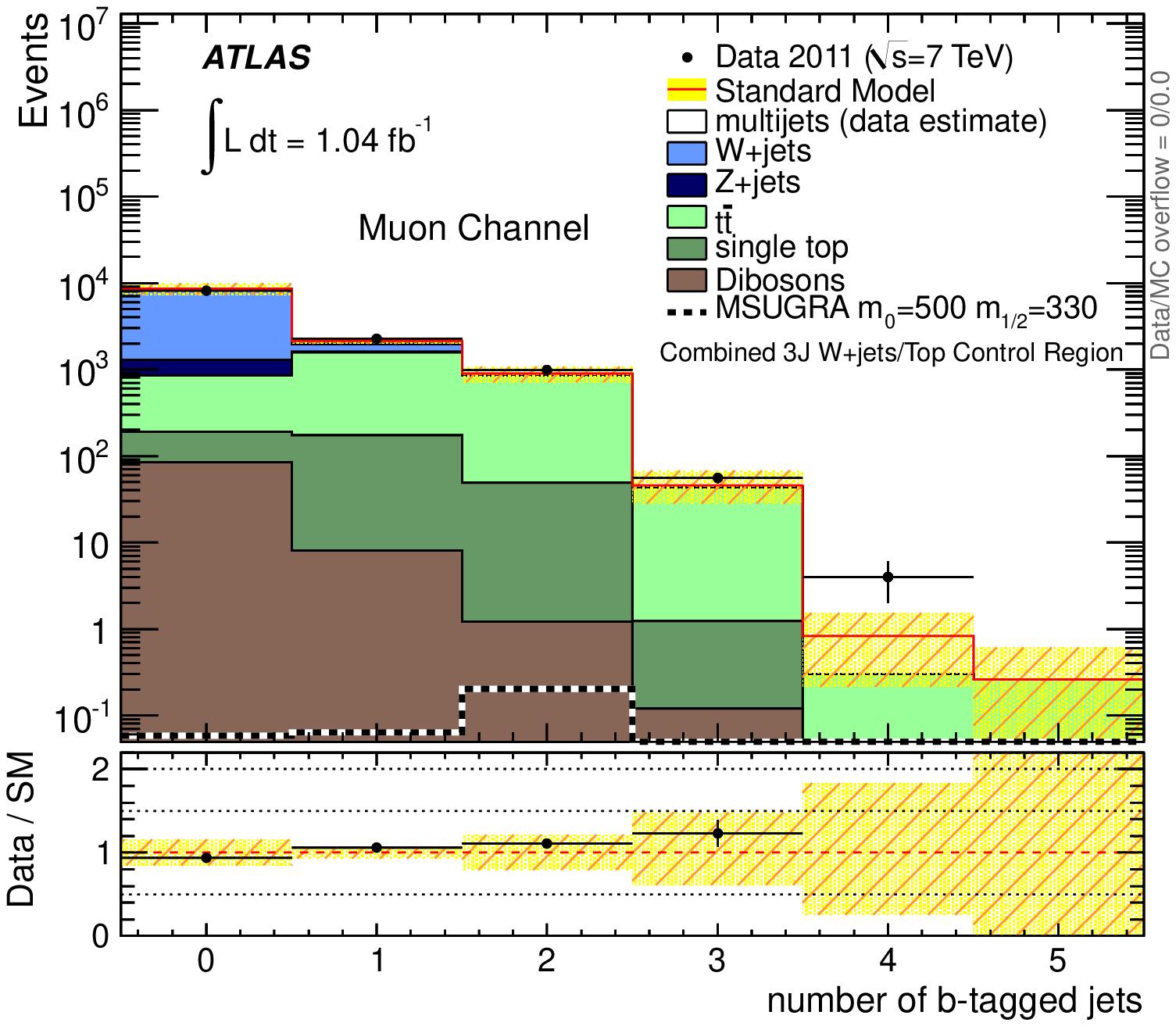}
\caption{
Distributions for events in the lepton plus three jets control regions 
for the electron
channel (left column) and muon channel (right column). Top row:
effective mass in the $W+$jets control region. Middle row: effective
mass in the top control region. Bottom row: number of $b$-tagged jets
in the combined $W+$jets and top control regions.
The ``Data/SM'' plots show the ratio between data and the summed 
standard model expectation.
The uncertainty band on the standard model expectation combines the MC statistical
uncertainty and systematic uncertainties on the jet energy scale and resolution,
$b$-tagging, the lepton resolution and identification efficiencies,
pile-up and luminosity.
}
\label{fig:lcontrol3}
\end{figure*}

\begin{figure*}
\center
\includegraphics[width=0.47\textwidth]{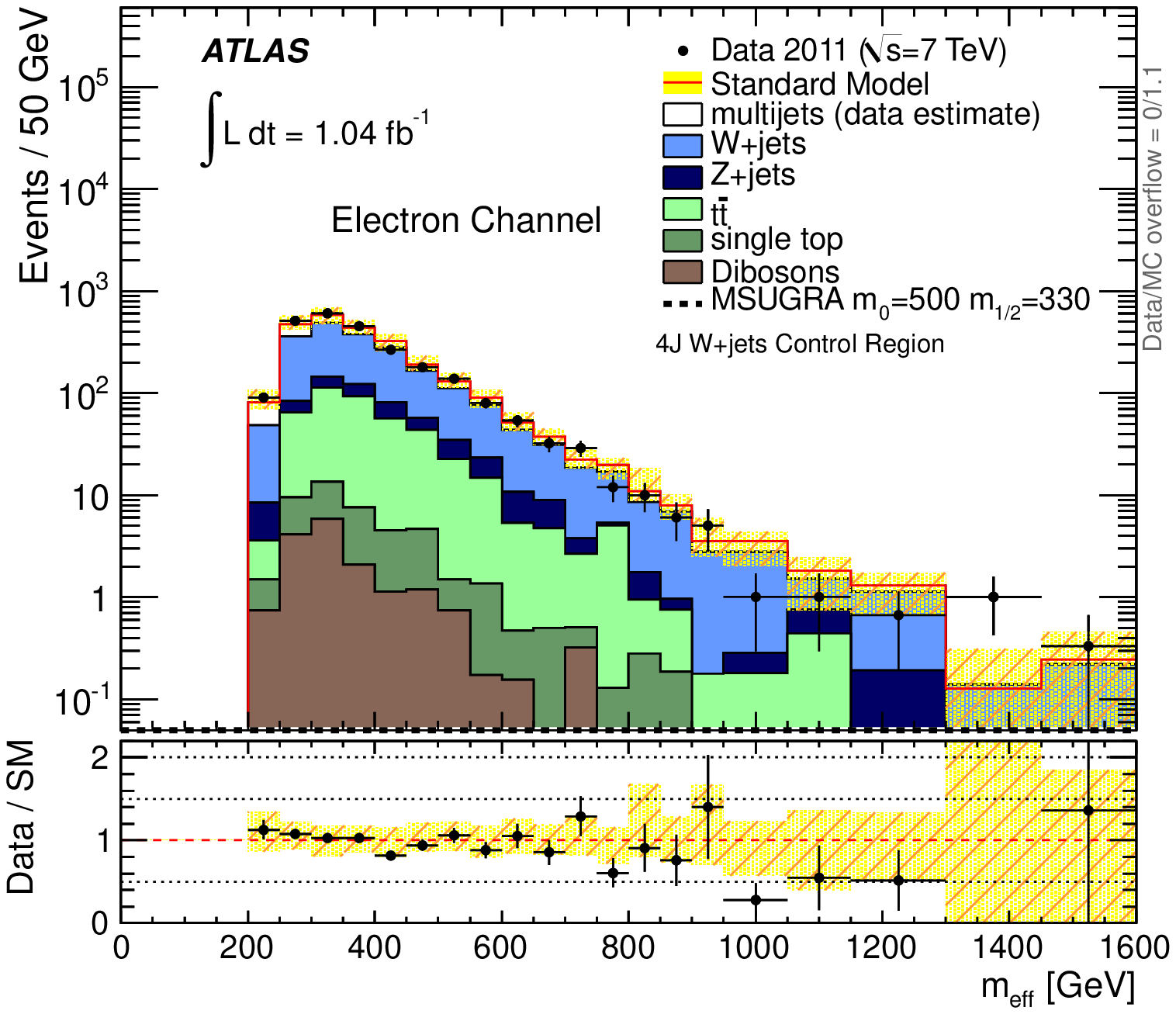}
\includegraphics[width=0.47\textwidth]{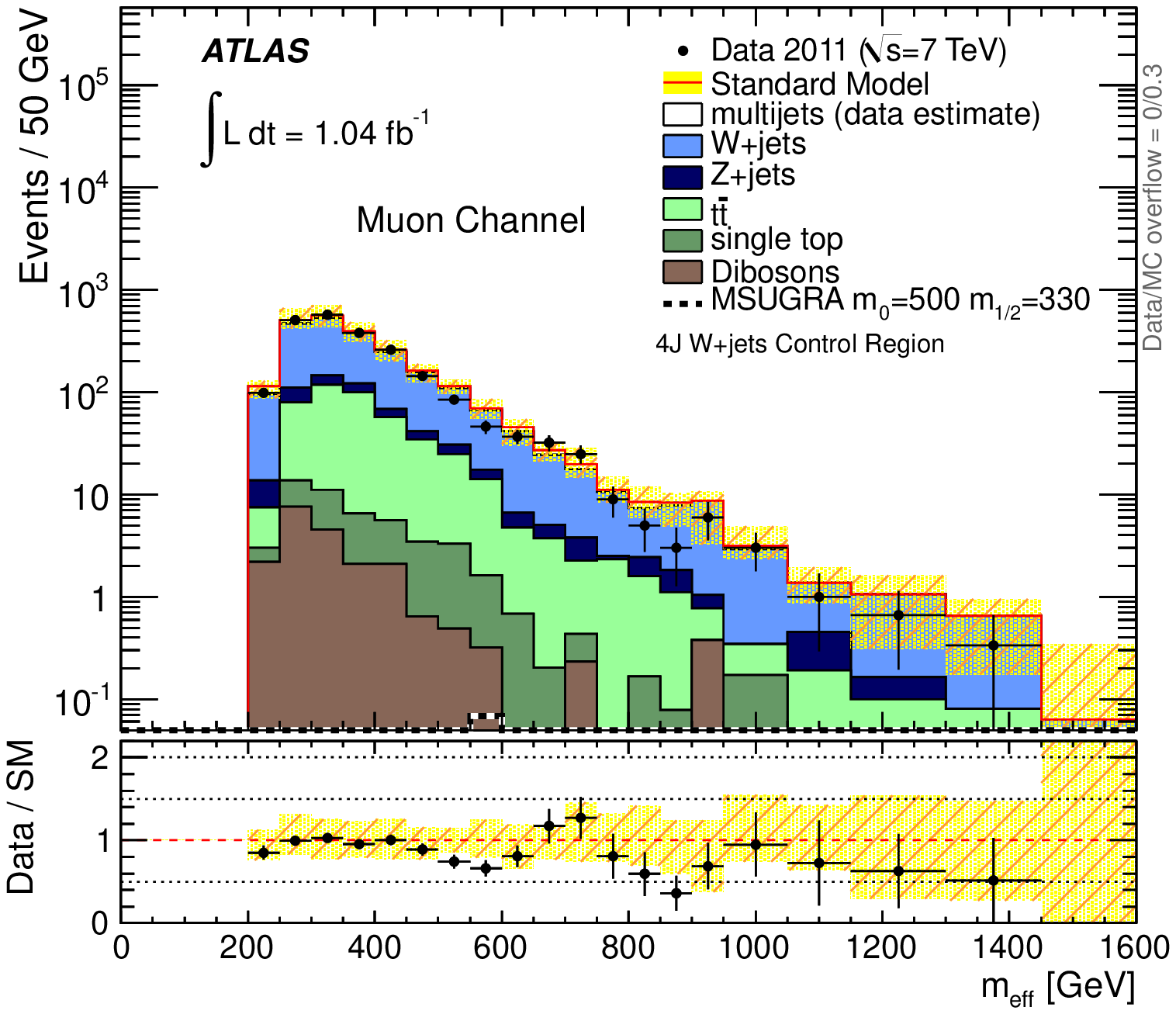}
\includegraphics[width=0.47\textwidth]{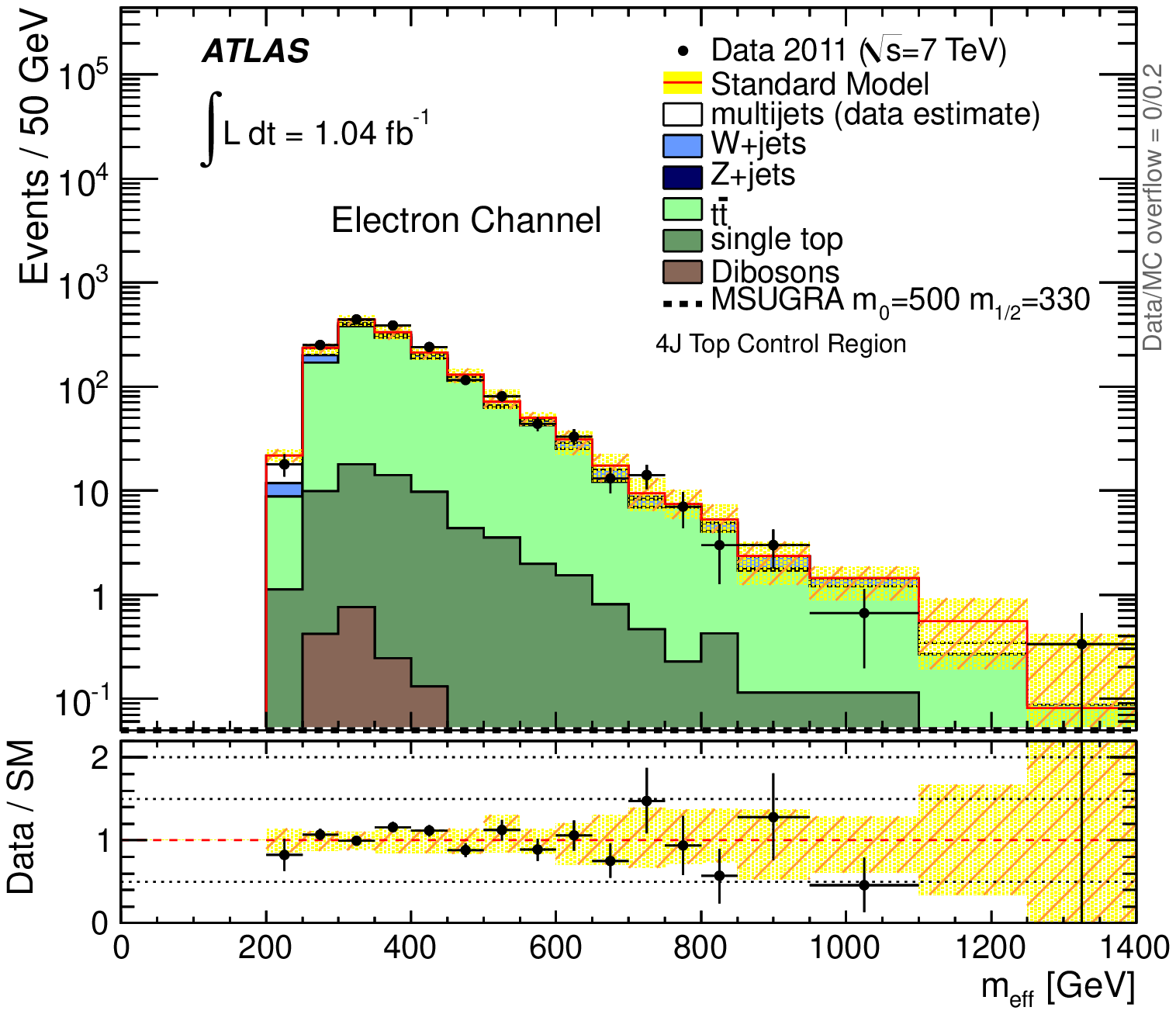}
\includegraphics[width=0.47\textwidth]{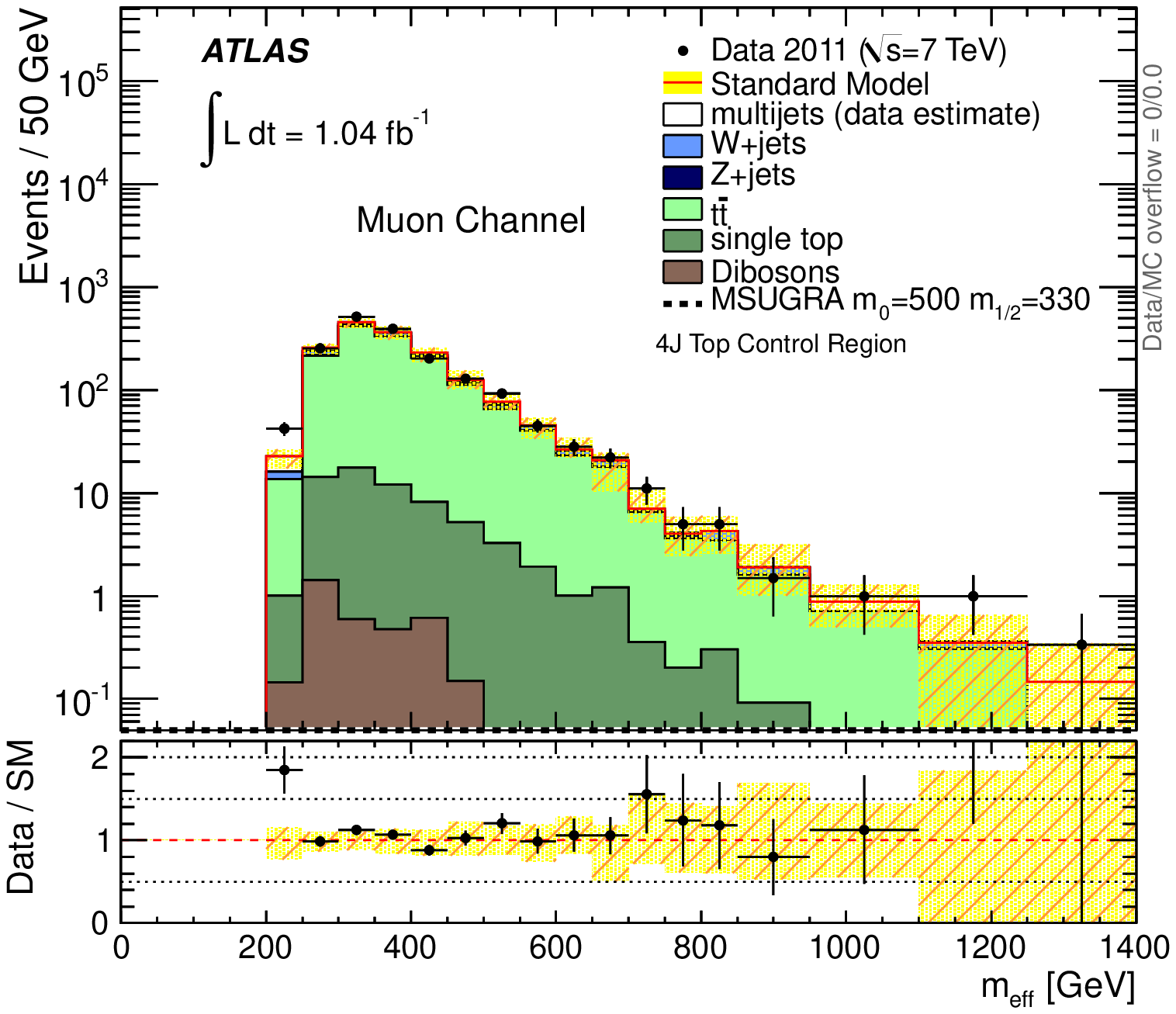}
\caption{
Distributions for events in the lepton plus four jets control regions 
for the electron
channel (left column) and muon channel (right column). Top row:
effective mass in the $W+$jets control region. Bottom row: effective
mass in the top control region. 
The ``Data/SM'' plots show the ratio between data and the summed 
standard model expectation.
The uncertainty band on the standard model expectation combines the MC statistical
uncertainty and systematic uncertainties on the jet energy scale and resolution,
$b$-tagging, the lepton resolution and identification efficiencies,
pile-up and luminosity.
}
\label{fig:lcontrol4}
\end{figure*}

\begin{enumerate}

\item {\it $W+$jets control regions} (WR).
$W+$jets control regions are defined by requiring $30 \GeV < \met
< 80 \GeV$, $40 \GeV < \mt < 80 \GeV$, and that none of the three or four
jets with the highest $\pt$ is tagged as a $b$-jet.

\item {\it Top control regions} (TR).
Top control regions are defined by identical cuts on $\met$ and
$\mt$ as for the $W+$jets control regions, but requiring at least one
$b$-tagged jet among the three or four 
jets with the highest $\pt$.

\end{enumerate}
The jet requirements are identical to the ones of the ``loose'' 
signal regions. In addition, a cut on $\meff$ is applied to both classes of control 
regions, again corresponding to the cut of the ``loose'' signal
regions: $\meff > 500 \GeV$ for the 3-jet selection, and 
$\meff > 300 \GeV$ for the 4-jet selection.
Figure~\ref{fig:lcontrol3} shows distributions
of $\meff$ and the number of $b$-tagged jets
for events in the $W+$jets and top control regions
for the electron and muon channel applying the 3-jet selection.
The distributions of $\meff$ in the 4-jet control regions are
shown in Fig.~\ref{fig:lcontrol4}.
The MC simulation describes the data well.

\section{Background Estimation}
\label{sec:bg}

The multijet background is estimated 
from the data in the signal regions and in the
$W+$jets and top control regions, using a matrix method. 
This background originates from
jets misidentified as leptons, but also from non-isolated real leptons,
for example from heavy flavor decay. In this paper, both
components are collectively called misidentified leptons.
For all regions, multijet-dominated samples are defined by
loosening the lepton identification criteria: 
for electrons the ``medium'' criteria are used instead of the ``tight'' 
criteria~\cite{electronefficiency}, and for
both electrons and muons the isolation criterion is dropped. 
Defining $N_{\mathrm{pass}}$ and $N_{\mathrm{fail}}$ as the number of events in such a loose
sample passing or failing the final lepton selection criteria,
and defining $N_{\mathrm{real}}$ and $N_{\mathrm{misid.}}$ as the number of real and
the number of misidentified leptons,
the following equations hold:
\begin{displaymath}
N_{\mathrm{pass}} = \epsilon_{\mathrm{real}} N_{\mathrm{real}} + 
\epsilon_{\mathrm{misid.}} N_{\mathrm{misid.}} ,
\end{displaymath}
\begin{displaymath}
N_{\mathrm{fail}} = (1 - \epsilon_{\mathrm{real}}) N_{\mathrm{real}} + 
( 1- \epsilon_{\mathrm{misid.}}) N_{\mathrm{misid.}} ,
\end{displaymath}
where $\epsilon_{\mathrm{real}}$ is the relative identification efficiency for
real leptons, and $\epsilon_{\mathrm{misid.}}$ is the misidentification
efficiency for misidentified leptons. Solving the equations leads to:
\begin{displaymath}
N_{\mathrm{misid.}}^{\mathrm{pass}} = \epsilon_{\mathrm{misid.}} N_{\mathrm{misid.}} =
\frac{N_{\mathrm{fail}} - (1/\epsilon_{\mathrm{real}} - 1) N_{\mathrm{pass}}}
{1/\epsilon_{\mathrm{misid.}} - 1/\epsilon_{\mathrm{real}}} .
\end{displaymath}
The efficiency $\epsilon_{\mathrm{real}}$ is taken from
simulated $Z \to ee$ events (electron channel) or $t \bar{t}$ and $W+$jets
events (muon channel). 
The efficiency $\epsilon_{\mathrm{misid.}}$ is determined from data
control samples enriched in multijet events, selected as follows. For the electron
channel ``medium'' electrons with $\pt > 20 \GeV$ are required. In
addition, one jet with $\pt > 30 \GeV$ needs to be present in the
event. To suppress $W$ and $t \bar{t}$ contributions, an upper cut
of 30 \GeV\ is imposed on \met. For the determination of the 
multijet background in the top CR, a $b$-tag is required for at
least one of the selected jets. For the muon final state, the multijet
control region is defined by one preselected muon with $\pt > 20
\GeV$, one jet with $\pt > 60 \GeV$ and $\met < 30 \GeV$. These 
control samples are corrected for contamination by real leptons,
which amounts to about 9\% for muons, and less than 3\% for electrons.
The misidentification efficiency $\epsilon_{\mathrm{misid.}}$ is measured
as function of $\pt$ and $\eta$ and this dependence is considered 
in the determination of the multijet contribution in both the signal and control 
regions. Typical values for $\epsilon_{\mathrm{real}}$ and
$\epsilon_{\mathrm{misid.}}$ are 88\% and 10\%, respectively, for the 
electron channel,
and 98\% and 35\%, respectively, for the muon channel.

A normalisation of the $W+$jets and top backgrounds to the data is
performed in the $W+$jets and top control regions.
Assuming that the shape of the distributions is described correctly
by the Monte Carlo simulation, transfer factors
$C^j_{i \mathrm{R} \to \mathrm{SR}}$ from control
region $i$R ($i=$W, T) to signal region SR for background type
$j$ ($j=$ $W+$jets, top) can be defined:
\begin{displaymath}
C^j_{i \mathrm{R} \to \mathrm{SR}} = \frac{N_{\mathrm{MC},j}^{\mathrm{SR}}}
{N_{\mathrm{MC},j}^{i \mathrm{R}}} .
\end{displaymath}
Thus the predicted contribution for background type $j$ in the signal region is
given by
\begin{displaymath}
N^{\mathrm{SR}}_{\mathrm{pred.} j} = \sum_{i=\mathrm{W,T}} (N^{i \mathrm{R}}_{\mathrm{data}} \times
C^j_{i \mathrm{R} \to \mathrm{SR}} ) .
\end{displaymath}
Typical values for the transfer factors are
$C^{\mathrm{W}}_{\mathrm{WR} \to \mathrm{SR}} = 0.023\ (0.007)$ and
$C^{t \bar{t}}_{\mathrm{TR} \to \mathrm{SR}} = 0.040\ (0.023)$ for the
electron channel and the 3JT (4JL) selection.
The control regions are not 100\% pure, and cross-contamination
of backgrounds in the various control regions is taken into
account. The solution of the coupled equations is
performed in a combined fit to each signal region and the
corresponding WR and TR control regions.
The estimated backgrounds include contributions from dileptonic events with
an undetected lepton as well as top quark or $W+$jets production with
leptonic tau decays.

The assumption that the MC simulation is able to predict the backgrounds
in the signal regions from the control regions is validated by checking additional
control regions at low $\mt$ and high $\met$, or at low $\met$ and high $\mt$.
Since these additional control regions have different kinematics and
composition than the nominal ones, 
these regions are susceptible to react differently to any mismodeling
of the data. In each region, the observed number of events is compared
to the prediction of the nominal background fit. In these 28
additional control regions, only one is found where the difference
between expected and observed events exceeds $2\sigma$.

Possible contamination from events originating from cosmic ray muons
is estimated by loosening the 
$|z_{\mu} - z_{\mathrm{PV}}| < 5$ mm requirement and studying the $z_{\mu}$
distribution, and is found 
to be negligible.
Remaining backgrounds from single top and diboson production
are estimated with MC simulation, and are also found to be negligible.

\section{Systematic uncertainties}
\label{sec:syst}

In this analysis systematic uncertainties arise on the estimates of
the background in the signal regions, as well as on the estimate
of the SUSY signal itself. The primary sources of
systematic uncertainty are the jet energy scale (JES) calibration,
the jet energy resolution (JER) uncertainty,
theory and MC modeling uncertainties,
and uncertainties on object reconstruction and identification.

The JES uncertainty has been measured from the complete 2010 data set 
using the techniques described in Ref.~\cite{jes}.
Additional contributions to the JES uncertainty are added to account for
the effect of pile-up at the relatively high luminosity delivered by the 
LHC in the 2011 run. The JES and JER calibrations are applied to MC simulated jets, and
their uncertainties are propagated throughout the analysis,
including to $\met$.

The JER measured with 2010 data~\cite{jer} is applied to all MC simulated 
jets. The difference in the JER between the re-calibrated and nominal MC 
simulation is taken as the systematic uncertainty.
Additional contributions are added to account for pile-up in 2011.

MC modeling uncertainties, affecting the transfer factors, are derived 
from alternative MC samples with different generators, or with 
different generator parameters.

Apart from jet energy scale, jet energy resolution and MC modeling
uncertainties,
further uncertainties on the background estimates originate from
finite MC statistics of top and $W+$jets events, from
lepton energy/momentum scale and resolution uncertainties,
from uncertainty in the lepton misidentification rates, from the
identification efficiencies for real leptons, and 
from $b$-tagging uncertainties.
The uncertainties on the background estimates are summarized in
Table~\ref{tab:bkgsyst}.


\begin{table*}
\caption{
Breakdown, in number of events, of the dominant systematic uncertainties on background 
estimates in the various signal regions.
Note that the nuisance parameters of individual uncertainties can be correlated in the fit,  
and therefore their uncertainties do not necessarily add quadratically to 
the total background uncertainty.
\label{tab:bkgsyst}}
\begin{center}
\setlength{\tabcolsep}{0.0pc}
\begin{tabular*}{\textwidth}{@{\extracolsep{\fill}}lcccc}
\noalign{\smallskip}\hline\noalign{\smallskip}
{\bf Electron channel}                        & 3JL & 3JT & 4JL & 4JT \\
\noalign{\smallskip}\hline\noalign{\smallskip}
Total statistical $(\sqrt{N_{\rm obs}})$ & $\pm 8.4$ & $\pm 3.7$  & $\pm 6.4$ & $\pm 3.0$ \\
Total background systematic              & $\pm 30.2$ & $\pm 7.4$  & $\pm 17.9$ & $\pm 3.7$ \\
\noalign{\smallskip}\hline\noalign{\smallskip}
\noalign{\smallskip}\hline\noalign{\smallskip}
Jet/$\met$ energy resolution    & $\pm 5.9$ & $\pm 0.5$  & $\pm 4.2$ & $\pm 0.8$ \\

Jet/$\met$ energy scale         & $\pm 18.6$ & $\pm 4.1$  & $\pm 13.6$ & $\pm 2.4$ \\
Lepton energy resolution        & $\pm 0.5$ & $\pm 0.3$  & $\pm 0.1$ & $\pm 0.3$ \\
Lepton energy scale             & $\pm 1.1$ & $\pm 0.3$  & $\pm 0.4$ & $\pm 0.5$ \\
$b$-tagging                     & $\pm 1.2$ & $\pm 0.2$  & $\pm 0.7$ & $\pm 0.1$ \\ 
MC stat. top                    & $\pm 5.8$ & $\pm 2.0$  & $\pm 3.8$ & $\pm 1.4$ \\

MC stat. W                      & $\pm 4.4$ & $\pm 2.3$  & $\pm 2.2$ & $\pm 1.3$ \\
Lepton misidentification rate   & $\pm 1.4$ & $\pm 0.1$ & $\pm 0.2$ & $< 0.1$ \\
Real lepton rate                & $\pm 1.5$ & $\pm 0.3$  & $\pm 0.8$ & $\pm 0.1$ \\
Top background modeling         & $\pm 15.9$ & $\pm 2.1$  & $\pm 9.8$ & $\pm 1.2$ \\
W background modeling           & $\pm 19.0$ & $\pm 5.6$  & $\pm 5.1$ & $\pm 1.9$ \\
Pile-up                         & $\pm 5.1$ & $\pm 1.0$  & $\pm 2.5$ & $\pm 0.4$ \\
\noalign{\smallskip}\hline\noalign{\smallskip}
\end{tabular*}
\begin{tabular*}{\textwidth}{@{\extracolsep{\fill}}lcccc}
{\bf Muon channel}                        & 3JL & 3JT & 4JL & 4JT \\
\noalign{\smallskip}\hline\noalign{\smallskip}
Total statistical $(\sqrt{N_{\rm obs}})$ & $\pm 7.6$ & $\pm 3.3$  & $\pm 7.1$ & $\pm 2.7$ \\
Total background systematic              & $\pm 19.3$ & $\pm 4.3$  & $\pm 15.8$ & $\pm 2.7$ \\
\noalign{\smallskip}\hline\noalign{\smallskip}
\noalign{\smallskip}\hline\noalign{\smallskip}
Jet/$\met$ energy resolution    & $\pm 9.0$ & $\pm 1.1$  & $\pm 0.9$ & $\pm 0.5$ \\

Jet/$\met$ energy scale         & $\pm 7.0$ & $\pm 0.2$  & $\pm 9.1$ & $\pm 1.6$ \\
Lepton energy resolution        & $< 0.1$ & $< 0.1$  & $< 0.1$ & $< 0.1$ \\
Lepton energy scale             & $\pm 0.8$ & $\pm 0.3$  & $\pm 1.4$ & $\pm 0.5$ \\
$b$-tagging                     & $\pm 1.0$ & $\pm 0.2$  & $\pm 0.9$ & $\pm 0.1$ \\
MC stat. top                    & $\pm 5.4$ & $\pm 2.1$  & $\pm 4.0$ & $\pm 1.4$ \\
MC stat. W                      & $\pm 2.5$ & $\pm 1.4$  & $\pm 2.6$ & $\pm 0.7$ \\
Lepton misidentification rate   & $< 0.1$ & $< 0.1$  & $< 0.1$ & $< 0.1$ \\
Real lepton rate                & $\pm 0.5$ & $\pm 0.1$  & $\pm 0.4$ & $< 0.1$ \\
Top background modeling         & $\pm 12.9$ & $\pm 2.4$  & $\pm 10.0$ & $\pm 1.2$ \\
W background modeling           & $\pm 8.8$ & $\pm 2.7$  & $\pm 7.3$ & $\pm 0.7$ \\
Pile-up                         & $\pm 3.5$ & $\pm 0.8$  & $\pm 2.7$ & $\pm 0.3$ \\
\noalign{\smallskip}\hline\noalign{\smallskip}
\end{tabular*}
\end{center}
\end{table*}
%

Systematic uncertainties on the SUSY signal are estimated through
variation of the factorisation and renormalisation scales in
PROSPINO between half and twice their default values, by considering
variations in $\alpha_s$, and by considering the PDF uncertainties
provided by CTEQ6. Uncertainties are calculated for individual
SUSY production processes. In the relevant regions of parameter space
in the MSUGRA/CMSSM model, these theoretical uncertainties on the
signal cross-sections are
typically $20-30$\%. Further uncertainties on the number of predicted
signal events arise from the JES uncertainty ($1-10$\%), the JER
uncertainty ($1-10$\%), pile-up uncertainties ($1-10$\%), lepton
trigger and identification uncertainties ($1-4$\%), the uncertainty on the
luminosity ($3.7$\%) and finite statistics of the signal Monte Carlo samples
($\sim 15$\%). Uncertainties in the modeling of initial state radiation
in signal events affect the uncertainty of the acceptance for low
values of squark and/or gluino masses, and for small mass differences in
the simplified models. These uncertainties are estimated from variations
of MC generator parameters as well as by explicitly generating
$\gluino \gluino +$jet and $\squark \squark +$jet events with a matrix
element approach as implemented in MadGraph 5~\cite{madgraph}.
Resulting uncertainties vary from negligible at high masses and high mass
splittings, to $\sim 30$\% at low masses and low mass splittings.

\section{Results and Interpretation}

\begin{figure*}
\center
\includegraphics[width=0.47\textwidth]{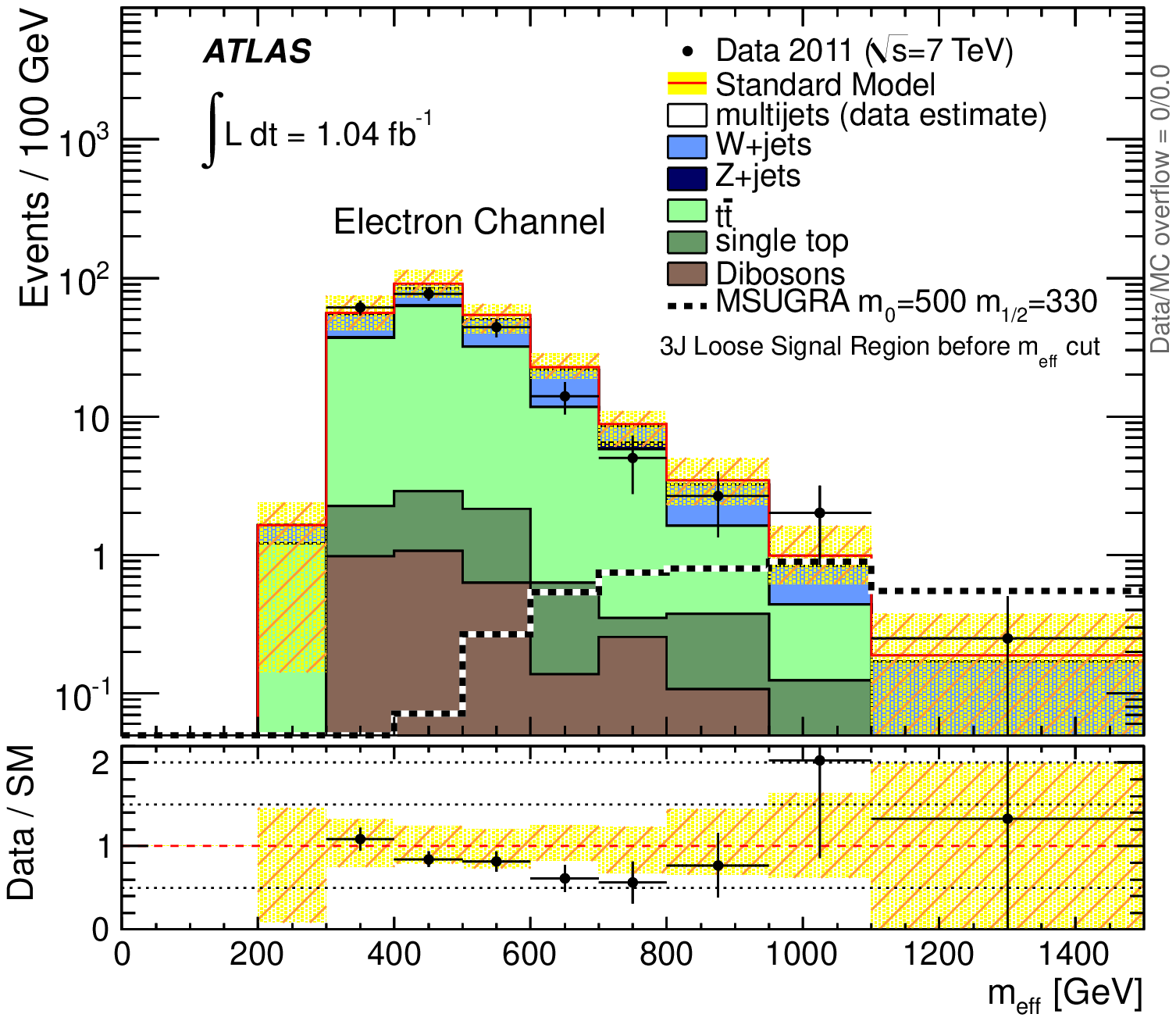}
\includegraphics[width=0.47\textwidth]{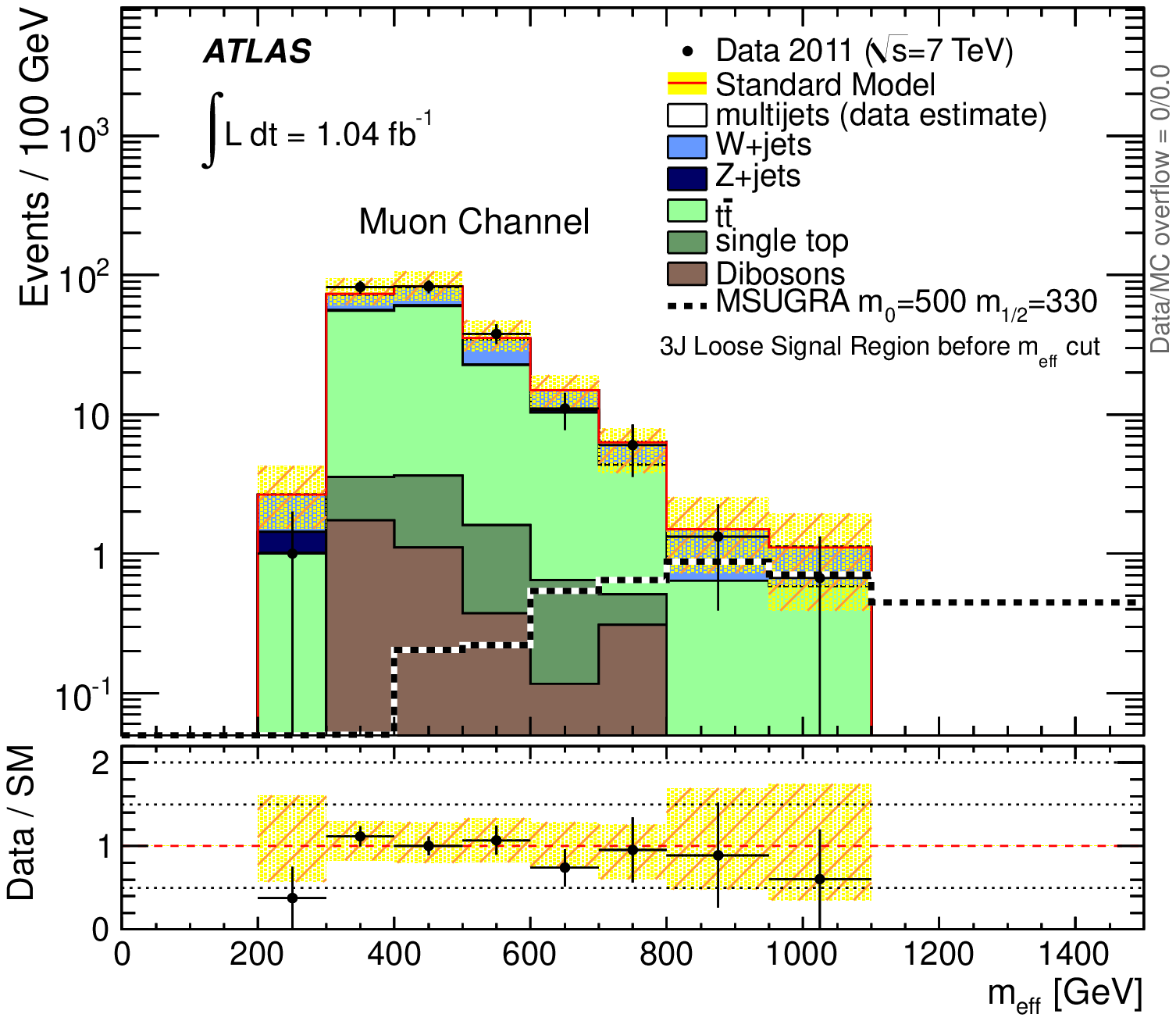}
\includegraphics[width=0.47\textwidth]{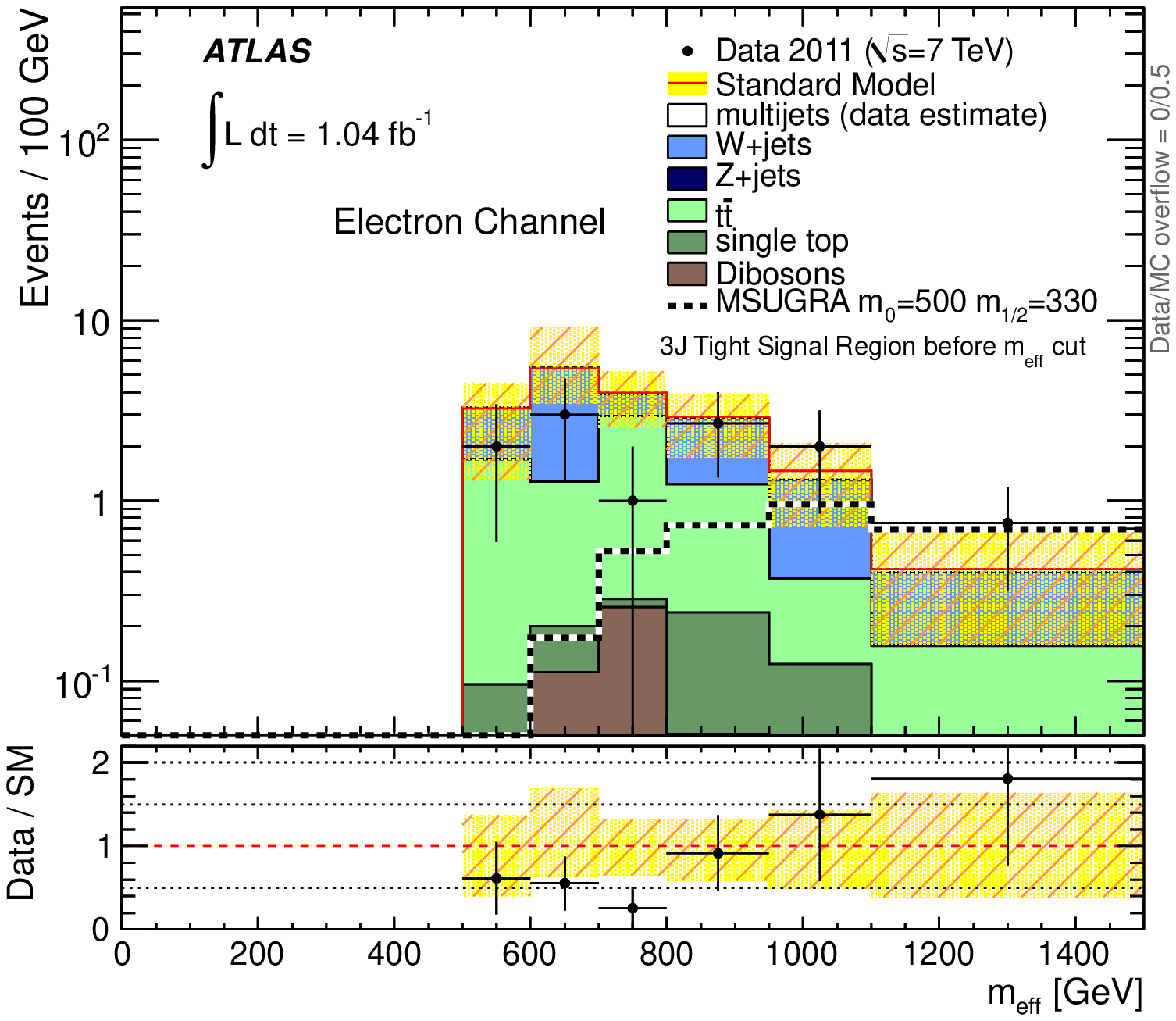}
\includegraphics[width=0.47\textwidth]{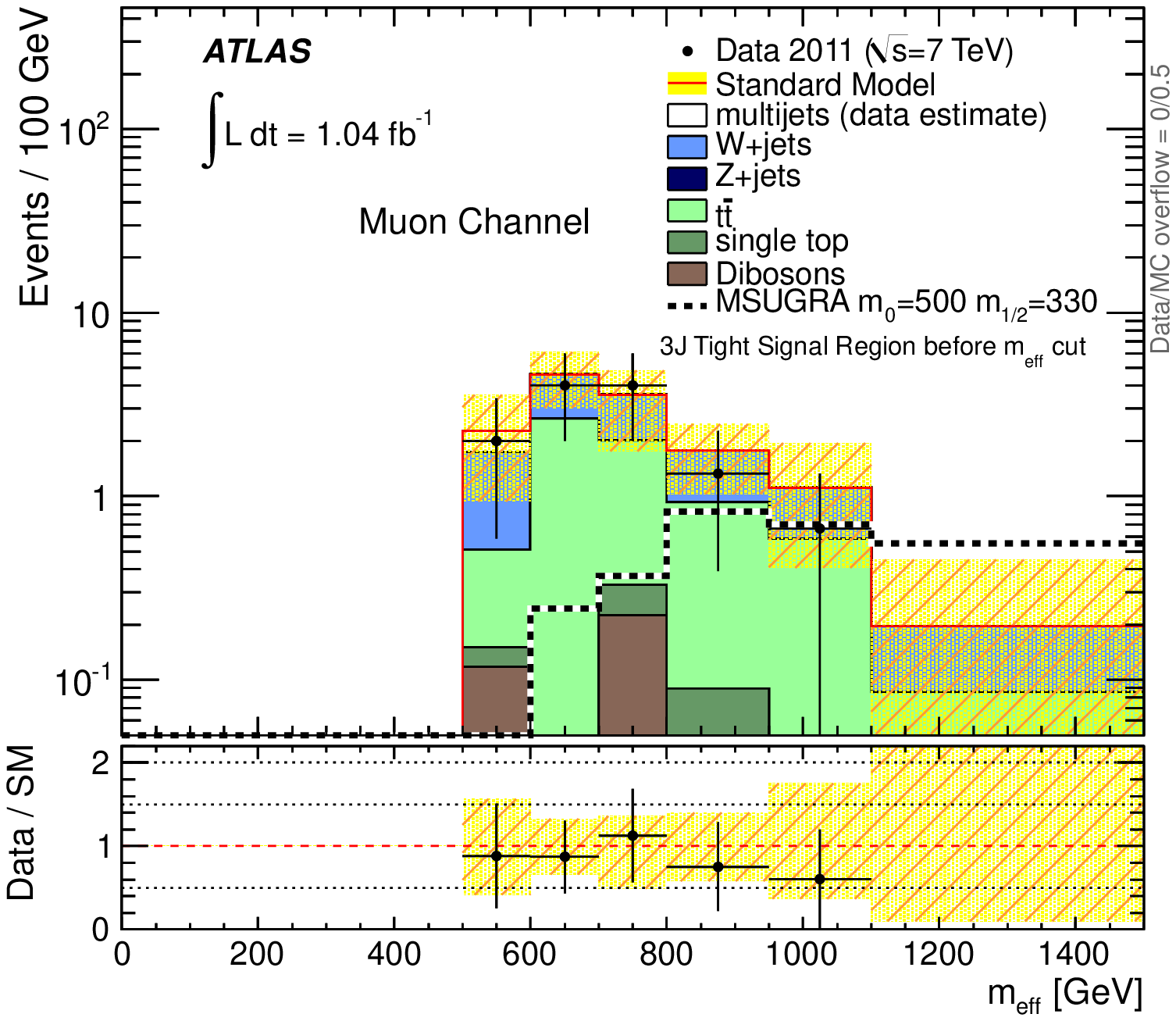}
\caption{
Distributions of the effective mass for events in the 3-jet signal 
regions 3JL (top) and 3JT (bottom) for
the electron channel (left) and the muon channel (right), after application 
of the final selection criteria described in Section~\ref{sec:sigregdef},
except for the cut on $\meff$ itself. 
The ``Data/SM'' plots show the ratio between data and the summed 
standard model expectation.
The uncertainty band on the standard model expectation combines the MC statistical
uncertainty and systematic uncertainties on the jet energy scale and
resolution, the lepton resolution and identification efficiencies,
pile-up and luminosity.
For illustration, the expected signal distributions of the MSUGRA/CMSSM
model point $m_0 = 500 \GeV$, $m_{1/2} = 330 \GeV$ are also shown.
}
\label{fig:lsignal3jet}
\end{figure*}

\begin{figure*}
\center
\includegraphics[width=0.47\textwidth]{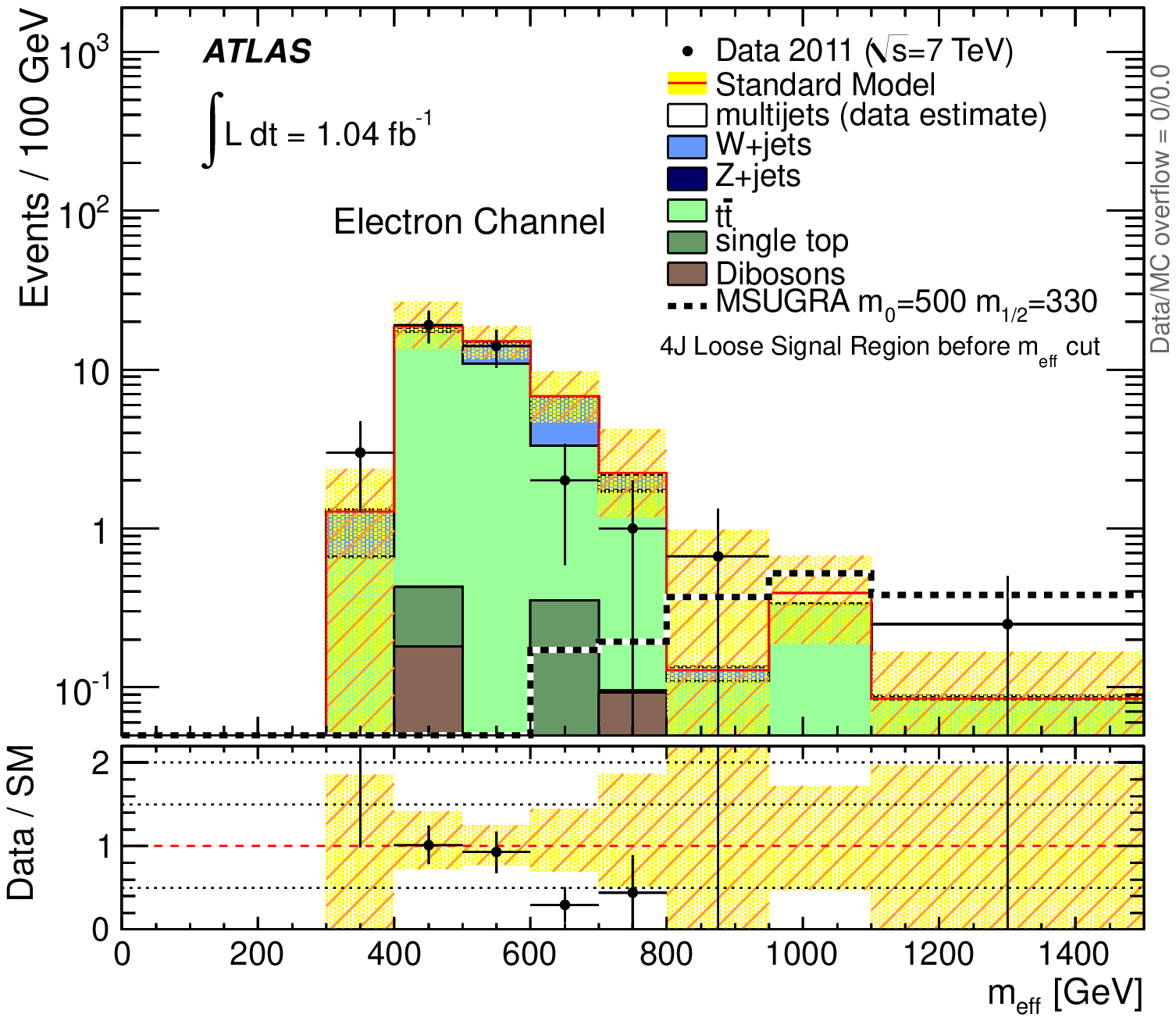}
\includegraphics[width=0.47\textwidth]{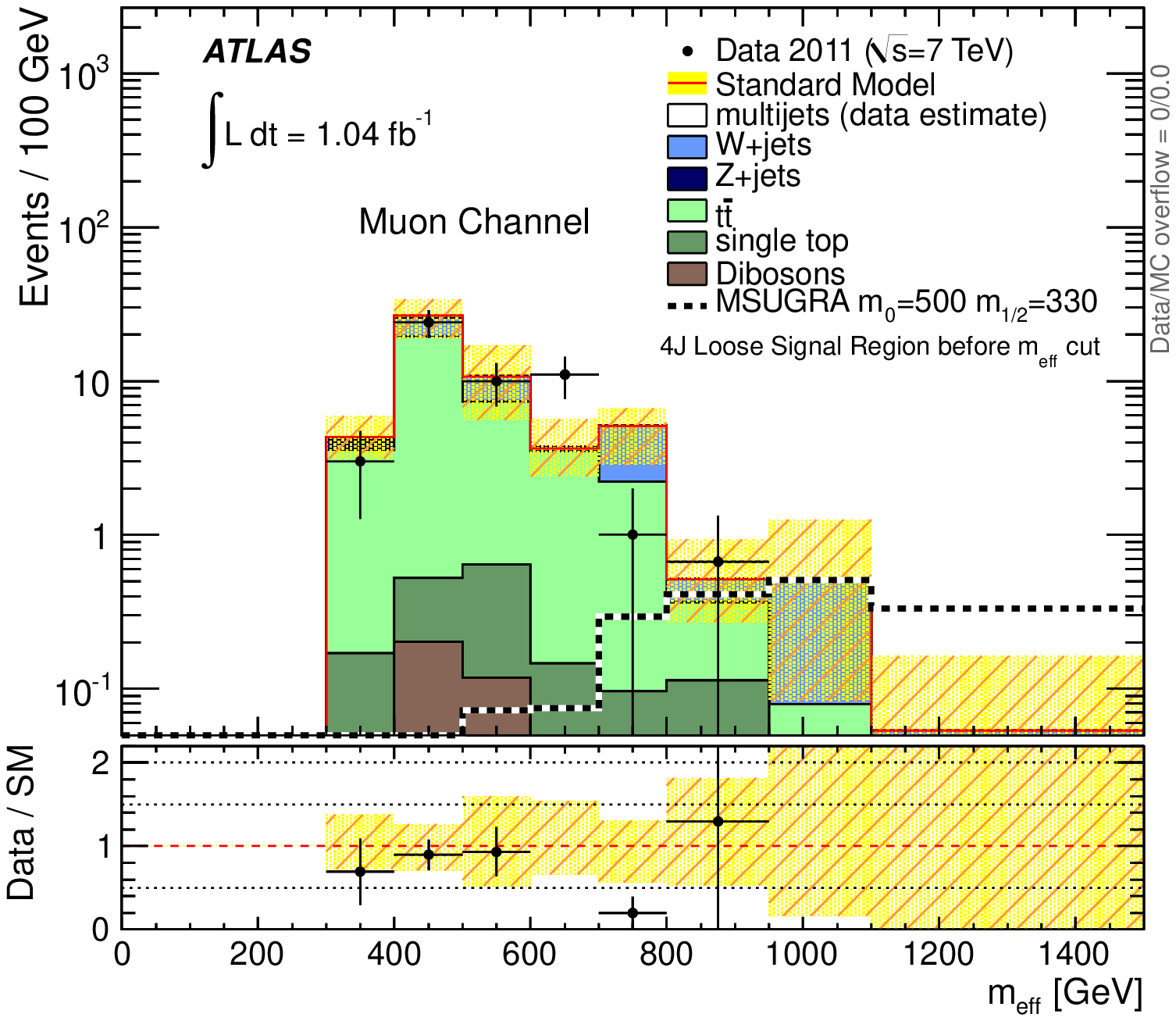}
\includegraphics[width=0.47\textwidth]{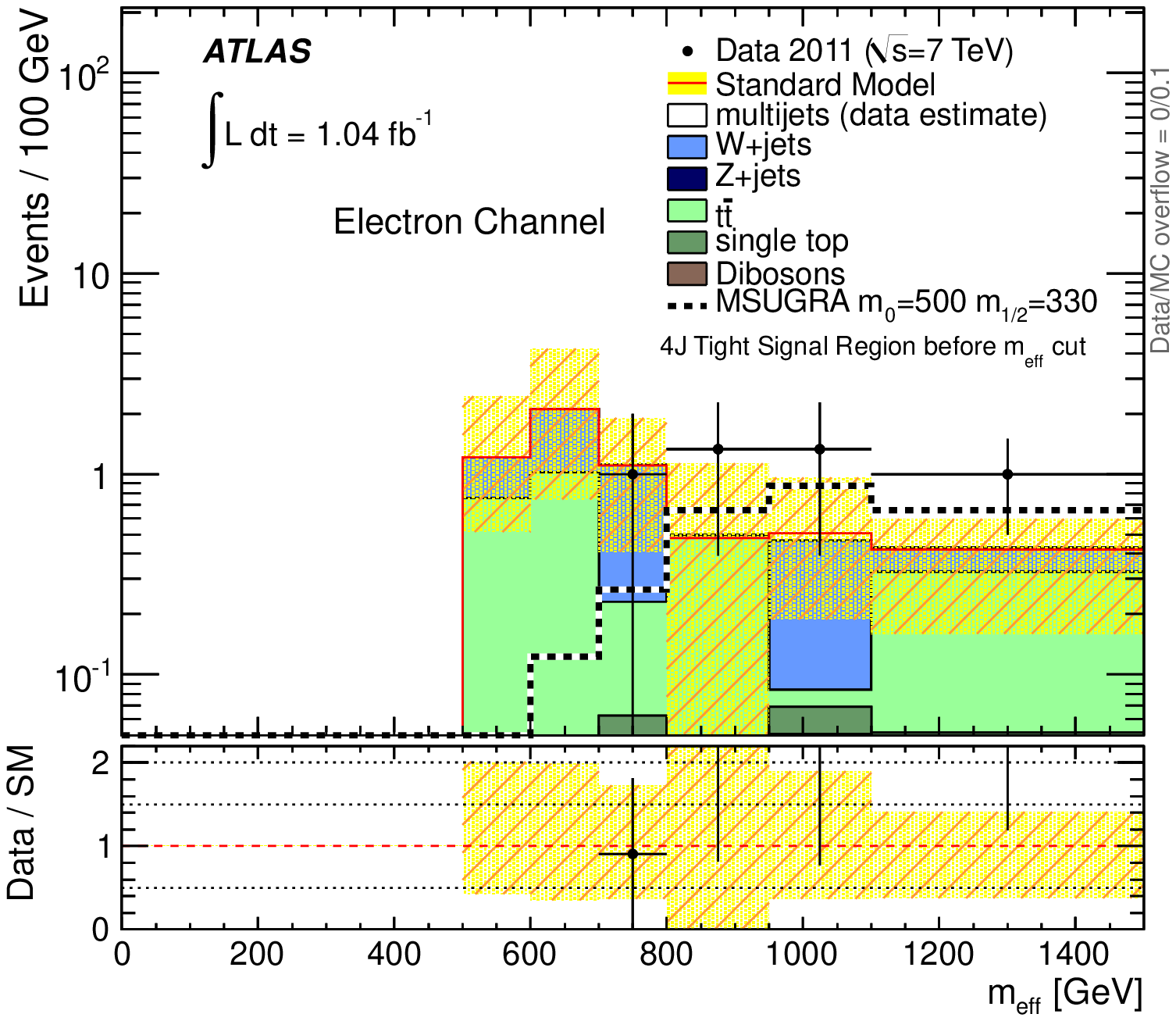}
\includegraphics[width=0.47\textwidth]{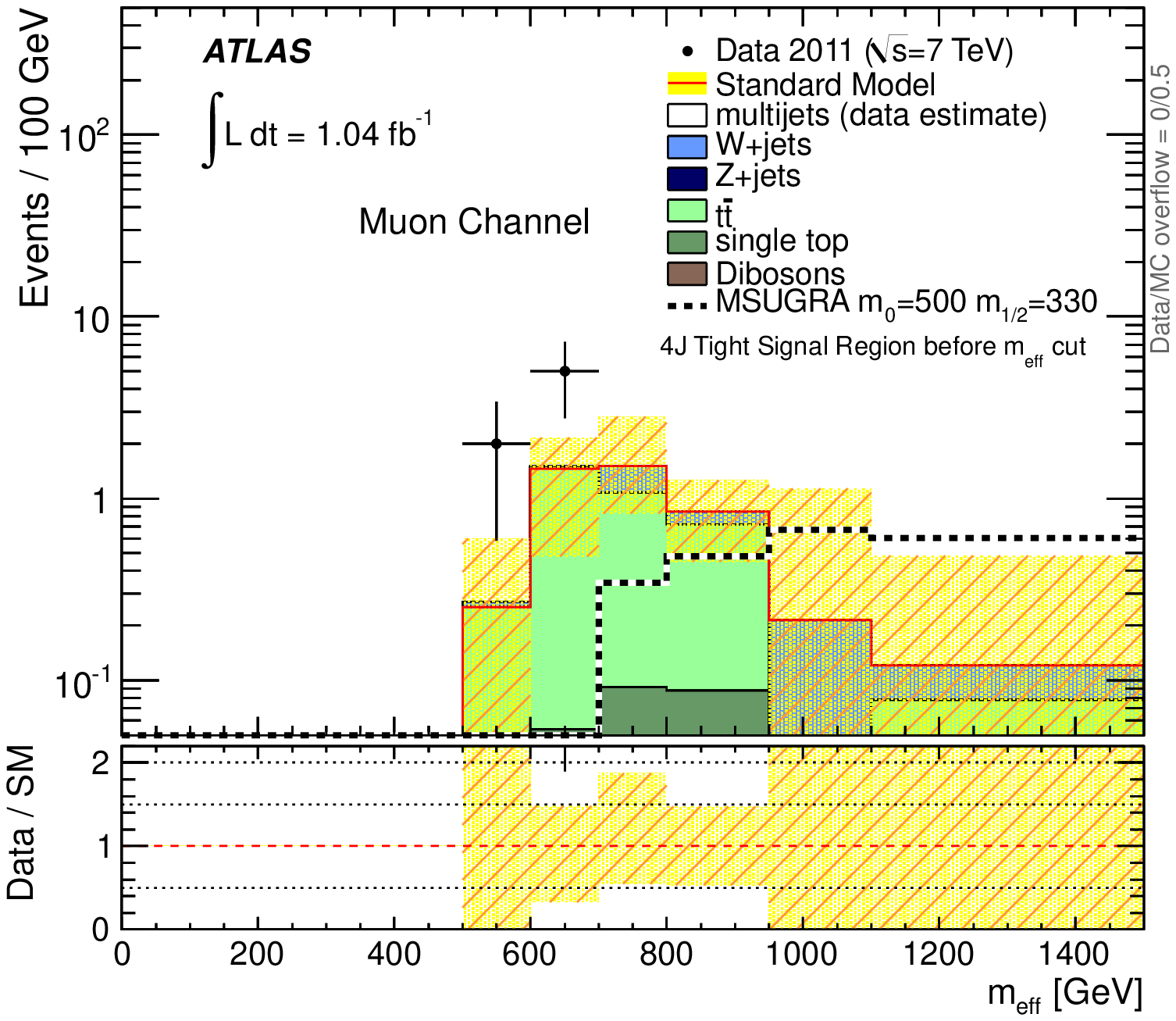}
\caption{
Distributions of the effective mass for events in the 4-jet signal 
regions 4JL (top) and 4JT (bottom) for
the electron channel (left) and the muon channel (right), after application 
of the final selection criteria described in Section~\ref{sec:sigregdef},
except for the cut on $\meff$ itself. 
The ``Data/SM'' plots show the ratio between data and the summed 
standard model expectation.
The uncertainty band on the standard model expectation combines the MC statistical
uncertainty and systematic uncertainties on the jet energy scale and resolution,
 the lepton resolution and identification efficiencies, pile-up and luminosity.
For illustration, the expected signal distributions of the MSUGRA/CMSSM
model point $m_0 = 500 \GeV$, $m_{1/2} = 330 \GeV$ are also shown.
}
\label{fig:lsignal4jet}
\end{figure*}

Figures~\ref{fig:lsignal3jet} and \ref{fig:lsignal4jet} show the distributions
of the effective mass in the 3-jet and 4-jet signal regions,
respectively, after application of the final selection criteria described in
Section~\ref{sec:sigregdef}, except for the cut on $\meff$
itself.


As discussed in Section~\ref{sec:bg}, a combined fit to the number 
of observed events in the signal and 
control regions is performed. 
The fit is performed for the four signal regions individually.
The likelihood function of the fit is written as:
\begin{displaymath}
L({\bf n}|s,{\bf b},{\btheta})  =  P_{\mathrm{S}} \,\times\,
P_{\mathrm{W}} \, \times \, P_{\mathrm{T}} \, \times \\
C_{\mathrm{Syst}}, 
\end{displaymath}
where ${\bf n}$ represents
the number of observed events in data, $s$ is the SUSY signal to be tested,
${\bf b}$ is the background, and ${\btheta}$ represents the systematic 
uncertainties, which are treated as
nuisance parameters with a Gaussian probability density function. 
The three $P$ functions in the right hand side
are Poisson probability distributions for event counts in the defined 
signal (S) and control regions (W and T, for $W$ and top pair,
respectively) 
and $C_{\mathrm{Syst}}$ represents the constraints on systematic
uncertainties.
Systematic uncertainties can be correlated between the signal and control regions.
The determination of the multijet contribution to the various regions, with
the method described in Section~\ref{sec:bg}, is performed as part 
of the fit procedure.

In ``discovery mode'', the number of SUSY signal events in the signal
regions is left free in the fit, as well as the background normalisations and nuisance parameters.
Possible signal contamination in the control regions is ignored. 
This fit tests the standard model hypothesis in the
signal regions, and quantifies any possible excess of events above the
background-only expectation in the signal regions.
The results of the ``discovery fit'' are shown in 
Tables~\ref{table:3jlt} and \ref{table:4jlt}.
Note that for the control regions, by construction, the number of 
``fitted'' background events equals the number of observed events.
The observed number
of events in data is consistent with the standard model expectation.
The last column in Table~\ref{tab:milimits} shows the $p-$values of
the discovery fit to data ($p(s=0)$ for the no-signal hypothesis) for the individual 
electron and muon channels.

\begin{table*}[tb]
\begin{center}
\setlength{\tabcolsep}{0.0pc}
\caption{
\label{table:3jlt}
Fit results for the electron (top part) and muon (bottom part) channels
in the ``loose'' 3-jet (3JL) and ``tight'' 3-jet (3JT) signal regions.
The results are obtained from the control regions using the 
``discovery fit'' (see text for details).
Nominal MC expectations (normalised to MC cross-sections) are given between
parentheses for comparison. 
}
{\footnotesize
\begin{tabular*}{\textwidth}{@{\extracolsep{\fill}}lrrrr}
\noalign{\smallskip}\hline\noalign{\smallskip}
 {\bf Electron channel} & 3JL Signal region & 3JT Signal region        & Top region              & $W$ region \\[-0.05cm]
\noalign{\smallskip}\hline\noalign{\smallskip}
Observed events     &  $71$                 & $14$                     & $162$                   & $565$ \\
\noalign{\smallskip}\hline\noalign{\smallskip}
Fitted top events   & $56 \pm 20$ ($51$)    & $7.6 \pm 3.0$ ($6.8$)    & $125 \pm 16$ ($112$)    & $64 \pm 8$ ($58$) \\
Fitted $W/Z$ events & $35 \pm 20$ ($34$)    & $10.5 \pm 6.5$ ($10.1$)  & $30.1 \pm 9.1$ ($29.3$) & $425 \pm 36$ ($413$) \\
Fitted multijet events   & $6.0^{+2.3}_{-1.4}$   & $0.46^{+0.37}_{-0.22}$   & $7.2 \pm 2.6$           & $76 \pm 24$ \\
\noalign{\smallskip}\hline\noalign{\smallskip}
Fitted sum of background events & $97\pm30$ & $18.5 \pm 7.4$           & $162 \pm 13$            & $565 \pm 24$ \\
\noalign{\smallskip}\hline\noalign{\smallskip}\noalign{\smallskip}
\end{tabular*}
\begin{tabular*}{\textwidth}{@{\extracolsep{\fill}}lrrrr}
\noalign{\smallskip}\hline\noalign{\smallskip}
 {\bf Muon channel}  &  3JL Signal region      & 3JT Signal region      & Top region              & $W$ region \\[-0.05cm]
\noalign{\smallskip}\hline\noalign{\smallskip}
Observed events      &  $58$                   & $11$                   & $166$                   & $413$ \\
\noalign{\smallskip}\hline\noalign{\smallskip}
Fitted top events    & $47 \pm 16$ ($38$)      & $8.9 \pm 3.2$ ($7.3$)  & $142 \pm 14$ ($115$)    & $70 \pm 7$ ($57$) \\
Fitted $W/Z$ events  & $16.6 \pm 9.4$ ($20.1$) & $5.0 \pm 3.2$ ($6.1$)  & $19.0 \pm 4.8$ ($23.2$) & $322 \pm 23$ ($393$) \\
Fitted multijet events    & $0.0 {}^{+0.0}_{-0.0}$  & $0.0 {}^{+0.6}_{-0.0}$ & $5.4 \pm 2.2$           & $21.6 \pm 5.7$ \\
\noalign{\smallskip}\hline\noalign{\smallskip}
Fitted sum of background events & $64\pm19$    & $13.9 \pm 4.3$         & $166 \pm 13$            & $413 \pm 20$ \\
\noalign{\smallskip}\hline\noalign{\smallskip}
\end{tabular*}}
\end{center}
\end{table*}

\begin{table*}[tb]
\begin{center}
\setlength{\tabcolsep}{0.0pc}
\caption{
\label{table:4jlt}
Fit results for the electron (top part) and muon (bottom part) channels
in the ``loose'' 4-jet (4JL) and ``tight'' 4-jet (4JT) signal regions.
The results are obtained from the control regions using the 
``discovery fit'' (see text for details).
Nominal MC expectations (normalised to MC cross-sections) are given between
parentheses for comparison. 
}
{\footnotesize
\begin{tabular*}{\textwidth}{@{\extracolsep{\fill}}lrrrr}
\noalign{\smallskip}\hline\noalign{\smallskip}
 {\bf Electron channel}  &  4JL Signal region     & 4JT Signal region      & Top region             & $W$ region \\[-0.05cm]
\noalign{\smallskip}\hline\noalign{\smallskip}
Observed events          &  $41$                  & $9$                    & $1382$                 & $1872$ \\
\noalign{\smallskip}\hline\noalign{\smallskip}
Fitted top events        & $38 \pm 15$ ($34$)     & $4.5 \pm 2.6$ ($4.1$)  & $1258 \pm 44$ ($1138$) & $391 \pm 14$ ($354$) \\
Fitted $W/Z$ events      & $9.5 \pm 7.5$ ($9.2$)  & $3.5 \pm 2.2$ ($3.4$)  & $88 \pm 21$ ($86$)     & $1242 \pm 89$ ($1202$) \\
Fitted multijet events        & $0.90^{+0.54}_{-0.37}$ & $0.00^{+0.02}_{-0.00}$ & $35 \pm 13$            & $239 \pm 78$ \\
\noalign{\smallskip}\hline\noalign{\smallskip}
Fitted sum of background events & $48 \pm 18$     & $8.0 \pm 3.7$          & $1382 \pm 37$          & $1872 \pm 43$ \\
\noalign{\smallskip}\hline\noalign{\smallskip}\noalign{\smallskip}
\end{tabular*}
\begin{tabular*}{\textwidth}{@{\extracolsep{\fill}}lrrrr}
\noalign{\smallskip}\hline\noalign{\smallskip}
 {\bf Muon channel}      &  4JL Signal region       & 4JT Signal region     & Top region              & $W$ region \\[-0.05cm]
\noalign{\smallskip}\hline\noalign{\smallskip}
Observed events          &  $50$                    & $7$                   & $1448$                  & $1623$ \\
\noalign{\smallskip}\hline\noalign{\smallskip}
Fitted top events        & $39 \pm 13$ ($36$)       & $4.7 \pm 2.2$ ($4.3$) & $1319 \pm 45$ ($1231$)  & $382 \pm 13$ ($357$) \\
Fitted $W/Z$ events      & $14.1 \pm 8.5$ ($14.2$)  & $1.4 \pm 1.1$ ($1.4$) & $91 \pm 19$ ($92$)      & $1169 \pm 46$ ($1185$) \\
Fitted multijet events        & $0.0 {}^{+0.0}_{-0.0}$   & $0.0^{+0.6}_{-0.0}$   & $38 \pm 10$             & $71 \pm 16$ \\
\noalign{\smallskip}\hline\noalign{\smallskip}
Fitted sum of background events & $53 \pm 16$       & $6.0 \pm 2.7$         & $1448 \pm 38$           & $1623 \pm 40$ \\
\noalign{\smallskip}\hline\noalign{\smallskip}
\end{tabular*}}
\end{center}
\end{table*}

Model-independent upper limits on new physics contributions to 
(only) the signal regions
can be derived from the discovery fit results.
The ignorance of possible signal contamination in the control
regions in the discovery fit leads to conservative upper limits on
non-standard model contributions.
The limits are derived using the $CL_s$ method~\cite{clsread}
based on the profile likelihood ratio test statistic~\cite{statforumlimits},
$\Lambda(s) = -2(\ln L({\bf n}|s,\hat{\hat{{\bf b}}},\hat{\hat{\btheta}}) -
\ln L({\bf n}|\hat{s},\hat{{\bf b}},\hat{\btheta}))$, where
$\hat{s}$, $\hat{{\bf b}}$ and $\hat{\btheta}$ maximize the likelihood
function and $\hat{\hat{{\bf b}}}$ and $\hat{\hat{\btheta}}$ maximize the likelihood for
a given choice of $s$. In the fit, $s$ and $\hat{s}$ are constrained to be non-negative.
The resulting 95\% confidence level (CL) limits are shown in
Table~\ref{tab:milimits} as observed and expected upper limits on the number of
non-SM events in the signal regions, as well as upper limits
on the visible cross-section (which equals the limit on the observed number
of signal events divided by the integrated luminosity).

\begin{table*}
\begin{center}
\caption{
\label{tab:milimits}
95\% CL upper limits on the visible cross-section ($\langle\epsilon\sigma\rangle^{95}_{\rm obs}$) and on the observed ($S^{95}_{\rm obs}$) and 
expected ($S^{95}_{\rm exp}$) number of signal events for the various signal regions. 
The last two columns indicate the $CL_B$ value and discovery $p$-value ($p(s=0)$). 
All numbers are given for the individual electron and muon channels.
}
\setlength{\tabcolsep}{0.0pc}
{\small
\begin{tabular*}{\textwidth}{@{\extracolsep{\fill}}lrrrrr}
\noalign{\smallskip}\hline\noalign{\smallskip}
{\bf Electron channel}  & $\langle\epsilon\sigma\rangle^{95}_{\rm obs}$ [fb] &   $S^{95}_{\rm obs}$    &    $S^{95}_{\rm exp}$    &    $CL_B$    &   $p(s=0)$   \\ 
\noalign{\smallskip}\hline\noalign{\smallskip}
 3JL  & $50$  & $52$      & $63^{+23}_{-11}$      & $0.21$ & $0.79$ \\
 3JT  & $14$  & $14.3$    & $16.5_{-3.0}^{+6.7}$  & $0.30$ & $0.71$ \\
 4JL  & $33$  & $34$      & $38^{+15}_{-7}$       & $0.35$ & $0.65$ \\
 4JT  & $10$  & $10.6$    & $9.5^{+4.3}_{-1.6}$   & $0.61$ & $0.42$ \\
\noalign{\smallskip}\hline\noalign{\smallskip}
{\bf Muon channel}  & $\langle\epsilon\sigma\rangle^{95}_{\rm obs}$ [fb] &   $S^{95}_{\rm obs}$    &    $S^{95}_{\rm exp}$    &    $CL_B$    &   $p(s=0)$   \\ 
\noalign{\smallskip}\hline\noalign{\smallskip}
 3JL  & $36$   & $38$    & $41^{+16}_{-7}$        & $0.39$ & $0.60$ \\ 
 3JT  & $10$   & $9.9$   & $11.4^{+4.5}_{-2.0}$   & $0.31$ & $0.70$ \\
 4JL  & $31$   & $32$    & $34^{+14}_{-7}$        & $0.42$ & $0.58$ \\
 4JT  & $9$    & $8.9$   & $8.0^{+3.0}_{-1.6}$    & $0.63$ & $0.39$ \\
\noalign{\smallskip}\hline\noalign{\smallskip}\noalign{\smallskip}
\end{tabular*}
}
\end{center}
\end{table*}

\begin{figure}
\center
\includegraphics[width=0.49\textwidth]{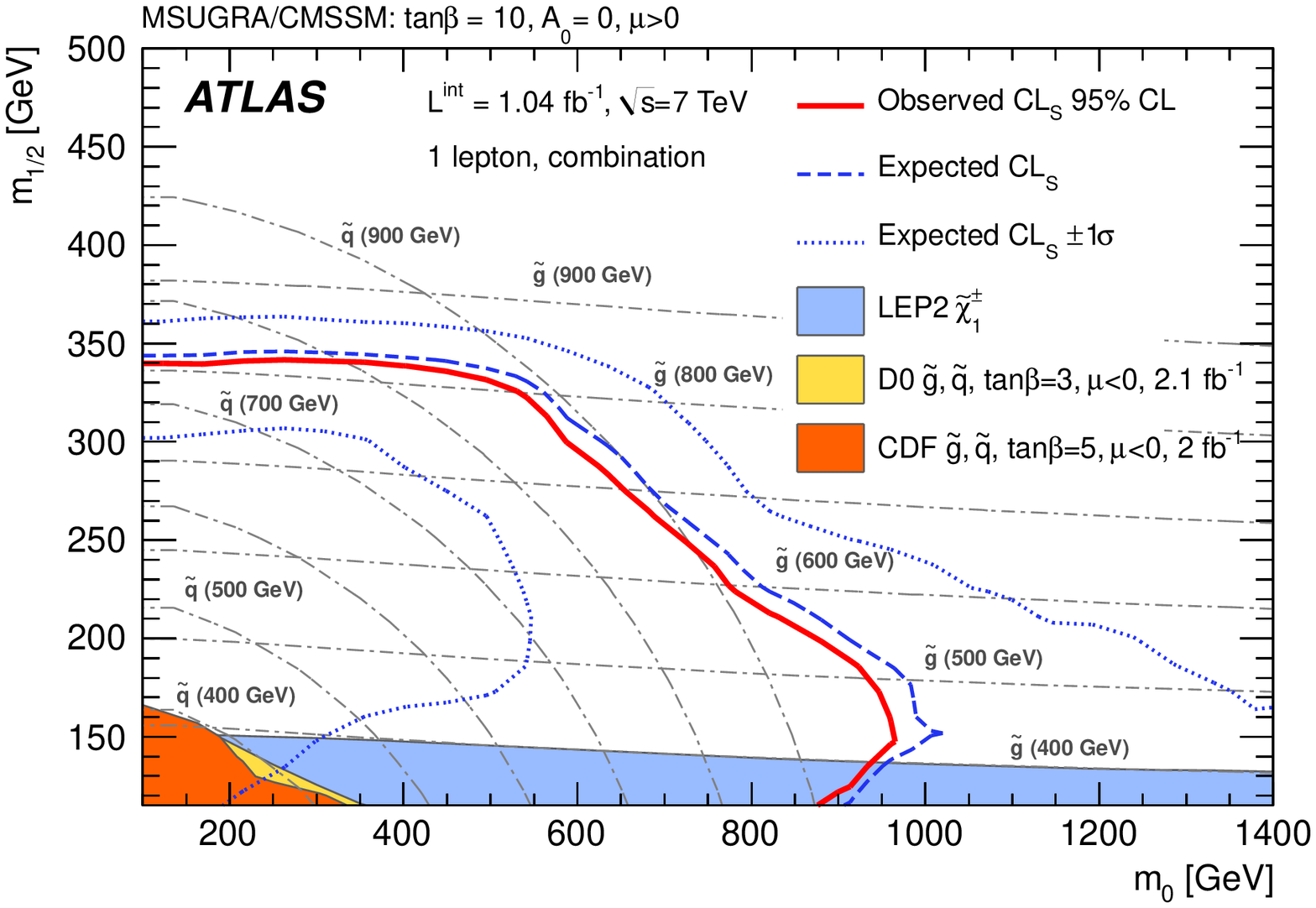}
\caption{Observed and expected 95\% CL exclusion limits, as well as the $\pm 1 \sigma$
variation on the median expected limit, in the combined
electron and muon channels.
The plots also show the published limits from
CDF~\cite{cdfsquarksgluinos},
D0~\cite{d0squarksgluinos}, and the results from the LEP 
experiments~\cite{lepsusy}.
}
\label{fig:cmssm}
\end{figure}

Limits within the MSUGRA/CMSSM framework are derived from a second fit to
signal and control regions, in ``exclusion mode''. This fit mode tests
for a specific new physics model, and uses signal predictions in the signal
regions as well as in the control regions.
The results are interpreted as limits for a grid of signal models in the 
$(m_0,m_{1/2})$ plane, as shown in Figure~\ref{fig:cmssm}. To combine the four
signal regions, the selection yielding the best expected limit for a given
parameter point is used.
The second-to-last column in Table~\ref{tab:milimits}
shows the values of $CL_B$, the confidence level for the background
hypothesis, which indicates the amount of
downward fluctuation of the observation, used in the 
$CL_s$ limit calculation.
Within the MSUGRA/CMSSM framework, and for equal squark and gluino masses,
gluino masses below $820 \GeV$
are excluded at 95\% CL by this analysis.
Varying $\tan \beta$ from $3$ to $10$, the limits are to a good
approximation independent of $\tan \beta$. For higher values of
$\tan \beta$, up to $\tan \beta = 40$, the effect on the limits 
depends on $m_0$ and $m_{1/2}$; for regions in the 
$(m_0,m_{1/2})$ plane with $m_{\squark} \approx m_{\gluino}$,
mass limits deteriorate by up to 10\%.

\begin{figure*}
\center
\includegraphics[width=0.47\textwidth]{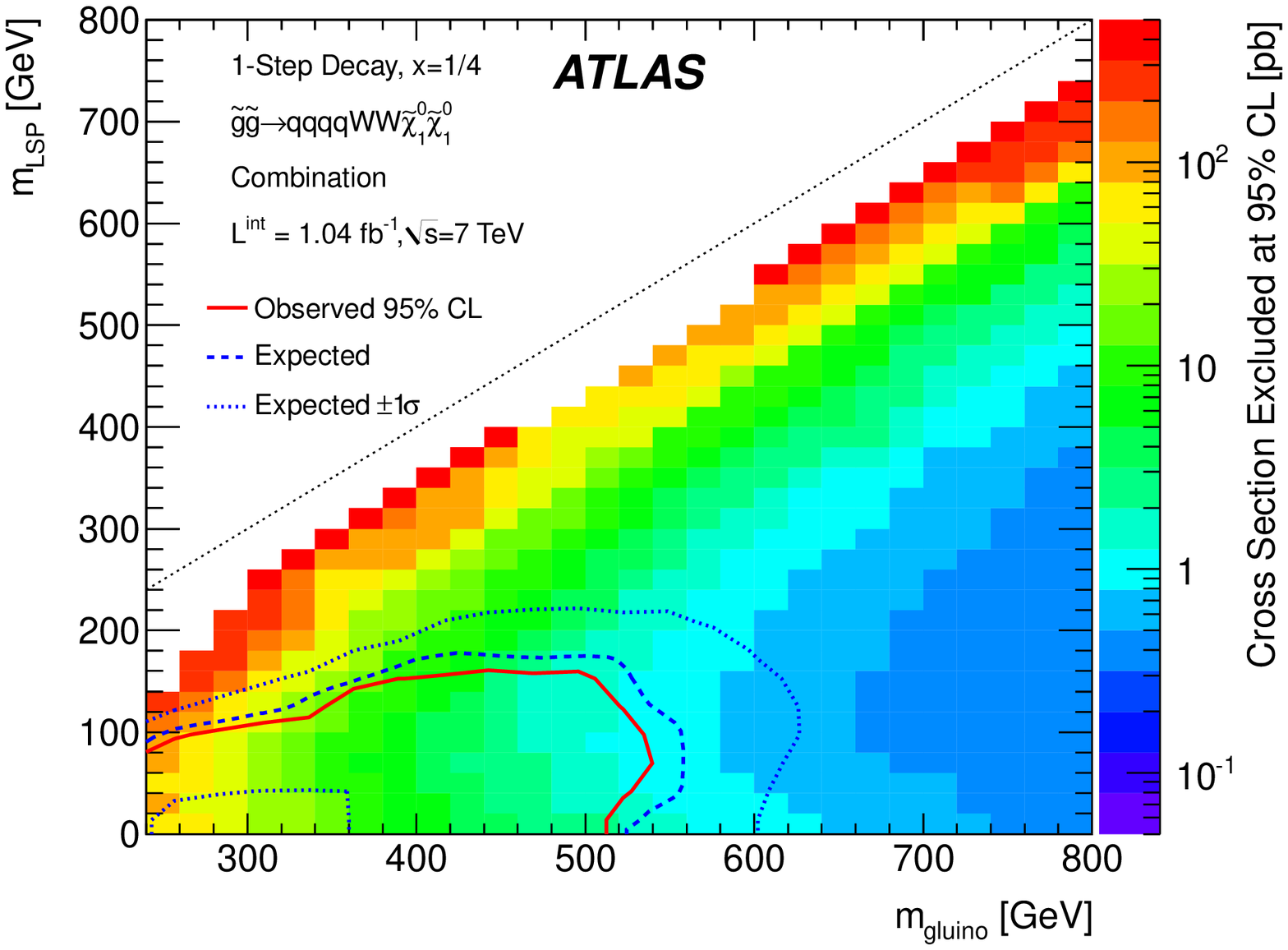}
\includegraphics[width=0.47\textwidth]{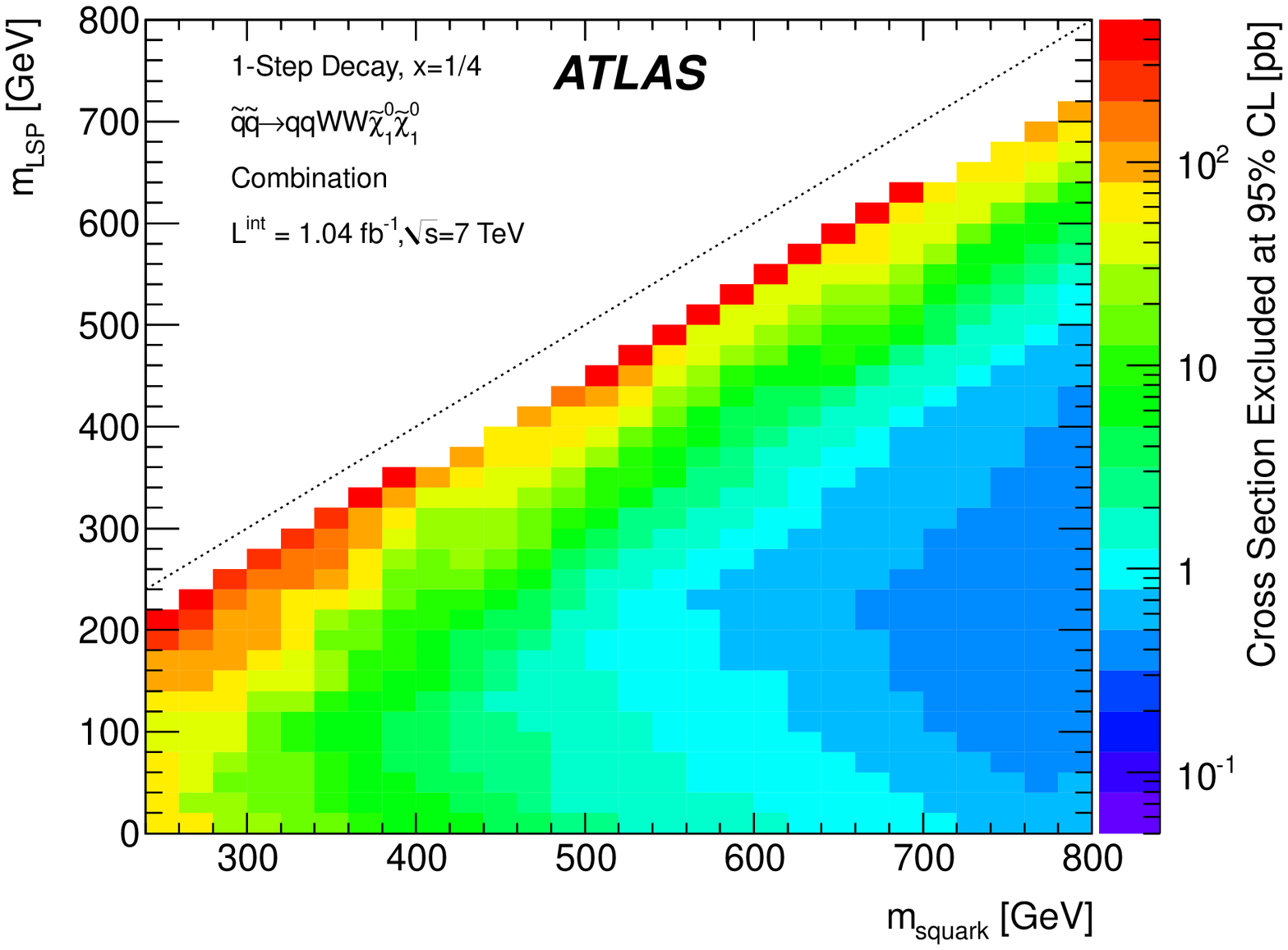}
\includegraphics[width=0.47\textwidth]{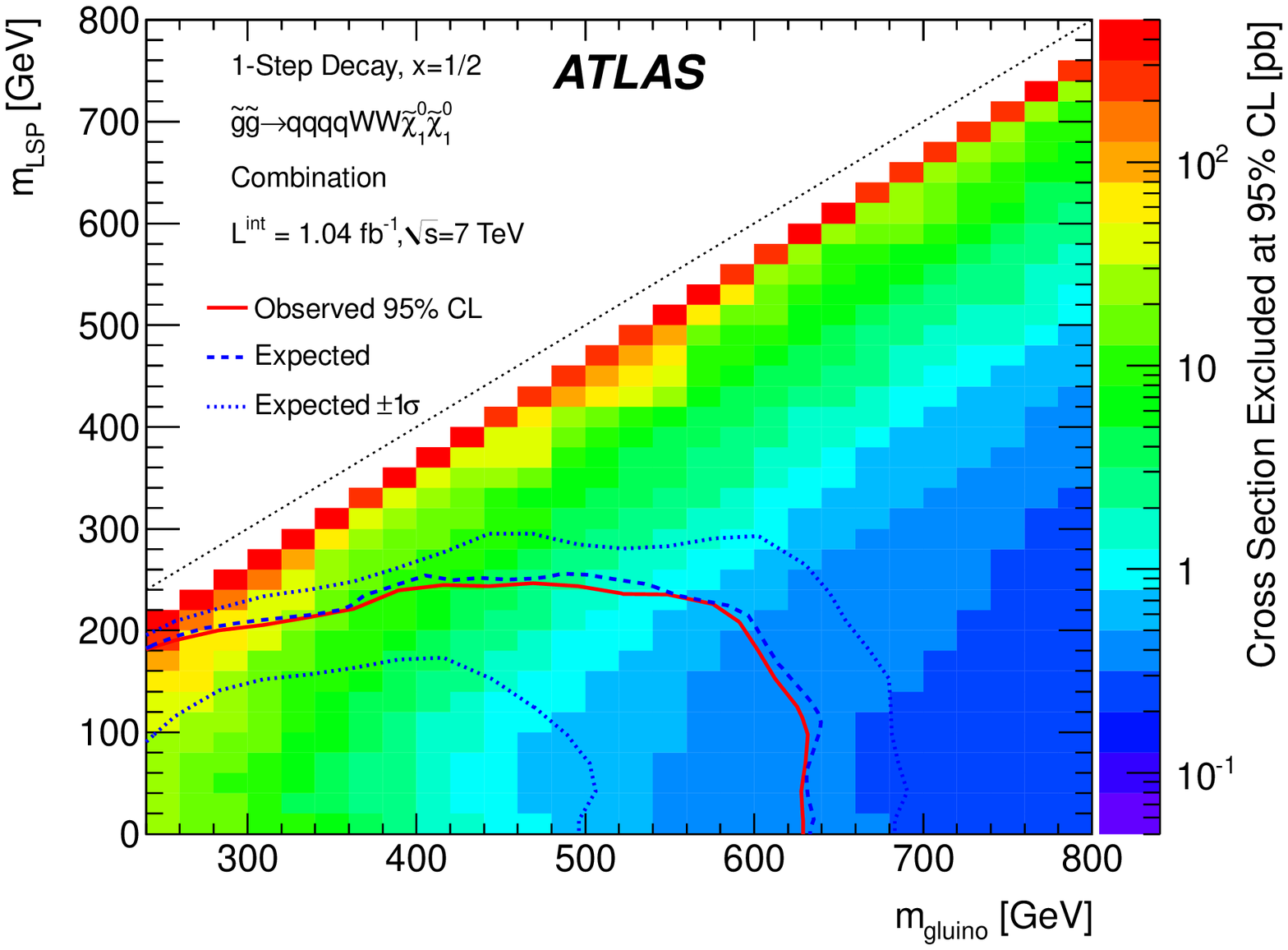}
\includegraphics[width=0.47\textwidth]{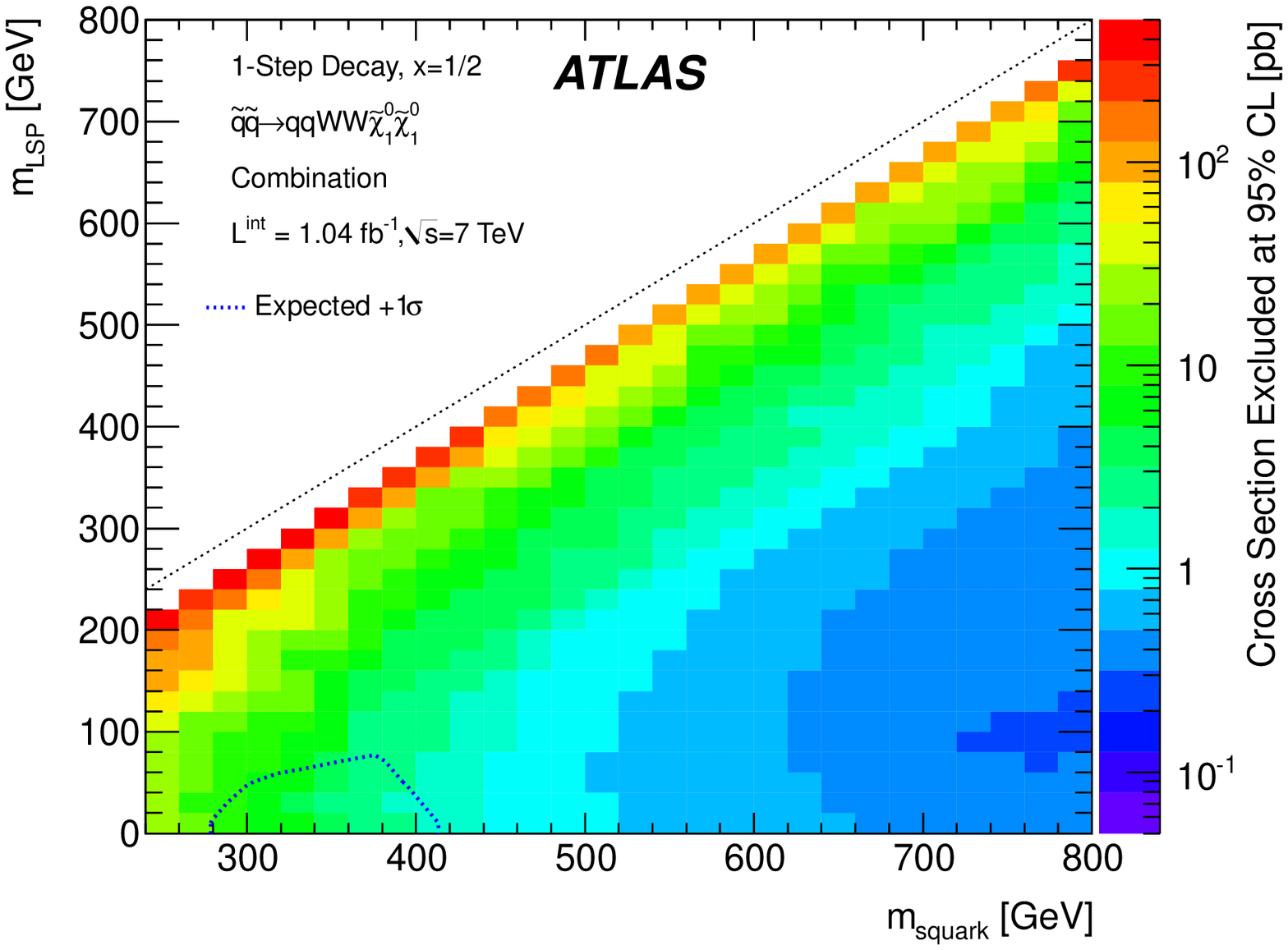}
\includegraphics[width=0.47\textwidth]{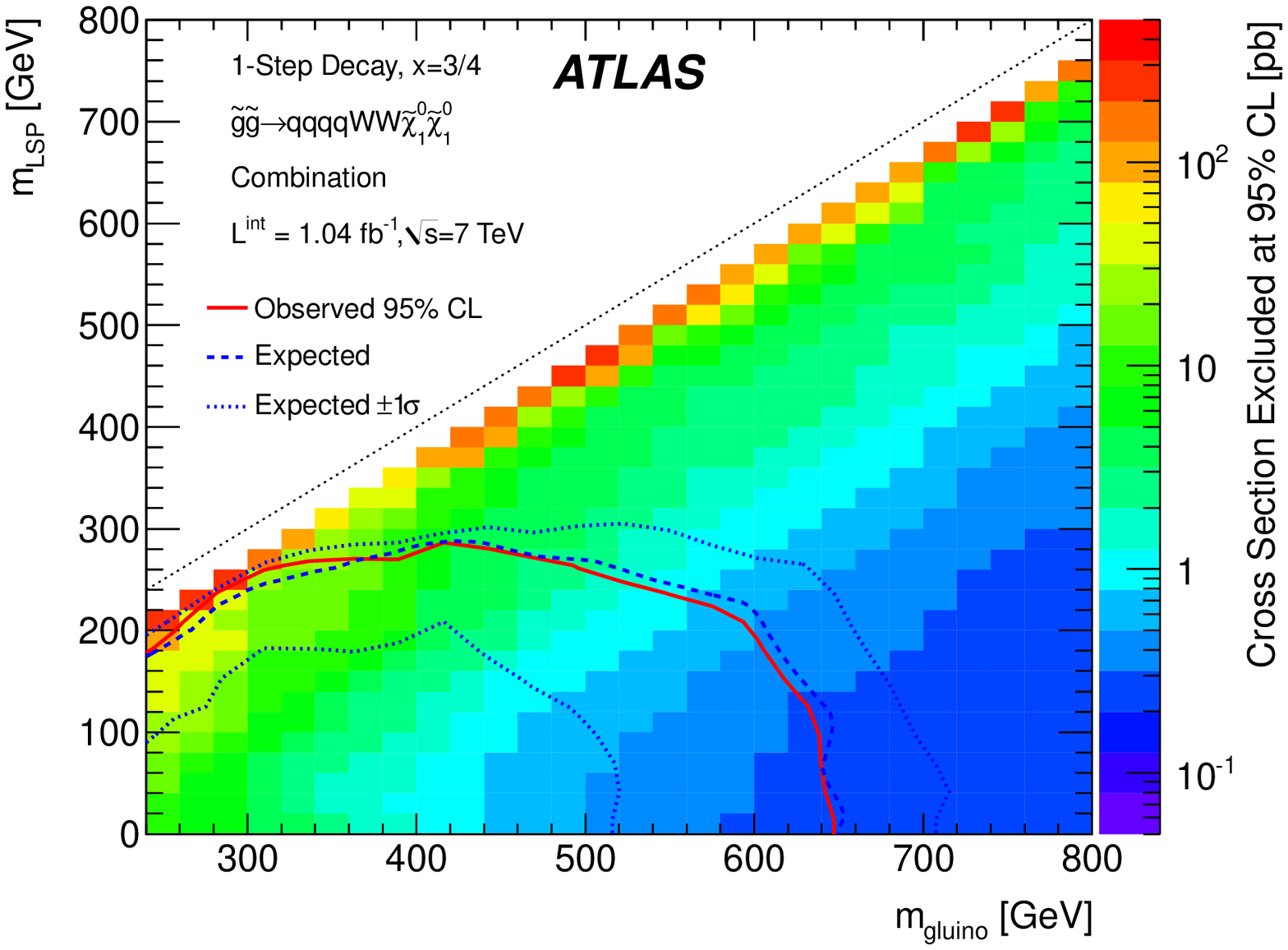}
\includegraphics[width=0.47\textwidth]{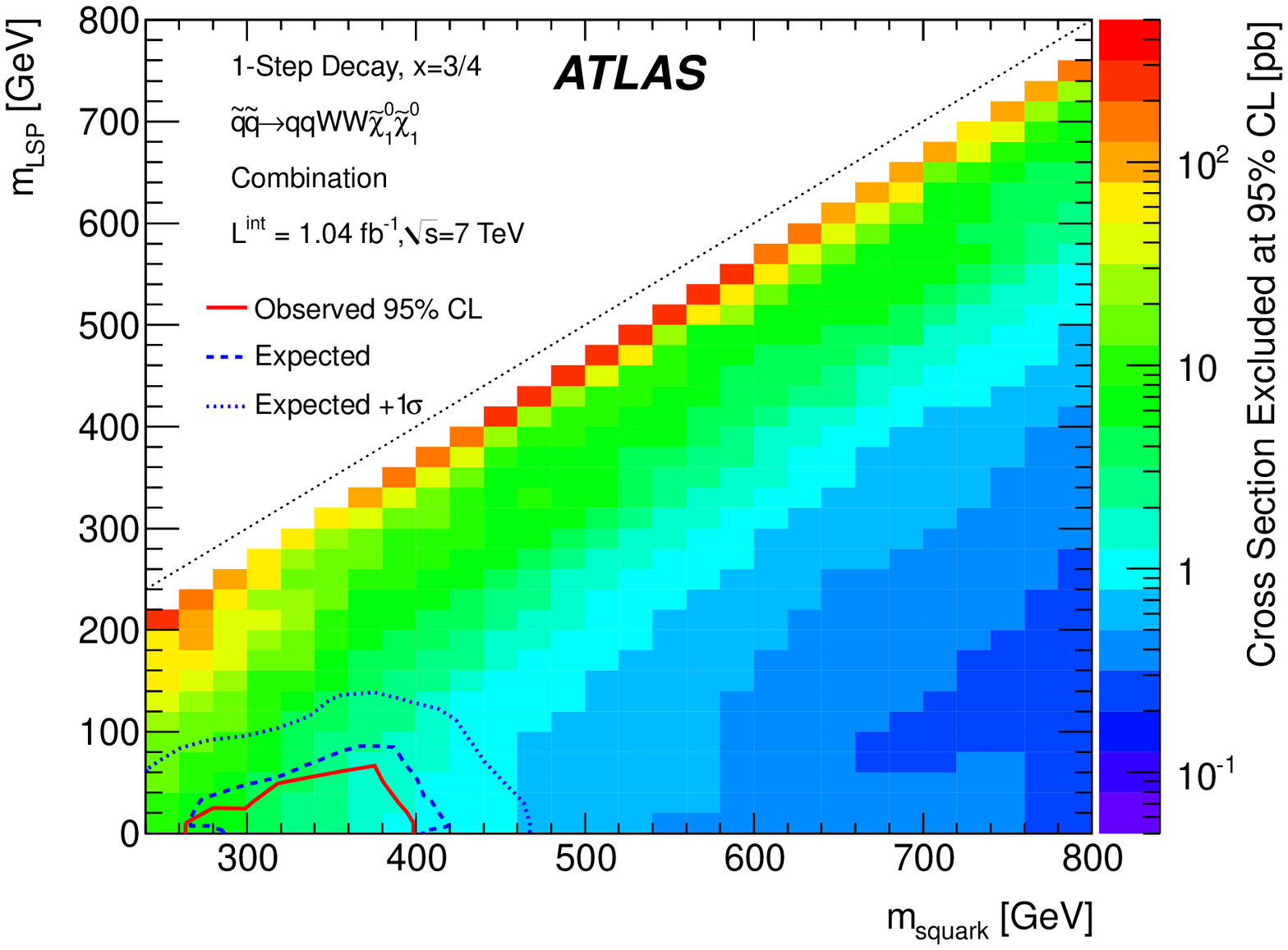}
\caption{Excluded cross-sections at 95\% confidence level for the
simplified models. The left column shows the results for the gluino models, the right
column shows the results for the squark models. The top row plots represent the case
$x = 1/4$, the middle row $x = 1/2$, and the bottom row $x = 3/4$.
The color coding (right axis) represents the model-independent cross-section limit.
Full lines indicate the observed exclusion regions in the shown
plane assuming production cross-sections as calculated with PROSPINO for the MSSM,
and a 100\% branching fraction into the assumed decay modes. The dashed
line shows the corresponding median expected limit and the dotted lines show the
$\pm 1 \sigma$ variation on the expected limit.}
\label{fig:simplified}
\end{figure*}

The results for the interpretation in terms of the simplified
models are shown in Figure~\ref{fig:simplified}. Again, the selection 
yielding the best expected limit for a given parameter point is used
for the combination of the four signal regions. The plots of
Figure~\ref{fig:simplified} show an upper limit on the
cross-section for new physics, at 95\% CL, as a function of
neutralino (LSP) and gluino or squark mass, for three different
values of the third free parameter, corresponding to the ratio 
of the mass differences in the relevant SUSY decay mode, 
$x = (m_{\chinopm} - m_{\nino})/
(m_{\gluino} - m_{\nino})$ (for the gluino models) or
$x = (m_{\chinopm} - m_{\nino})/ (m_{\squark} - m_{\nino})$ (for
the squark models). To obtain these upper limits,
identical cross-sections are assumed for the electron and muon channels, 
and no theoretical uncertainties are considered.
The plots of Figure~\ref{fig:simplified} show that the limits on the
cross-section for
new physics deteriorate when the LSP mass approaches
the squark or gluino mass, i.e. when the mass spectrum is compressed.
Also indicated on the plots are the observed
exclusion regions, assuming production cross-sections
as calculated with PROSPINO for the MSSM, and a 100\% branching
fraction into the assumed decay modes. 
In the gluino model, all squark masses are set to $4.5 \TeV$ and
only gluino pair production is considered.
In the squark model, the masses
of the gluino and of the third generation squarks are set to $4.5
\TeV$. The masses of the left- and right-handed squarks of the first
and second generation are set to be equal.
By setting the gluino mass to $4.5 \TeV$, the t-channel (gluino exchange) 
production of $\squarkL \squarkR$ is effectively suppressed.
In supersymmetric theories such as the MSSM only the left-handed squarks 
decay to charginos with 100\% wino content, which is implied by this
particular simplified model.
Therefore the PROSPINO
squark pair production cross section is divided by a factor two to
obtain the $\squarkL \squarkL$ cross section.
Note that reducing the gluino mass to $1.2 \TeV$ would increase
this cross section by a few percent for $m_{\squark} = 200 \GeV$,
but by a factor two for $m_{\squark} = 400 \GeV$.
For the calculation of the
exclusion regions, theoretical uncertainties on the cross-sections,
as discussed in Section~\ref{sec:syst}, are taken into account.
In the gluino model at high $x$,
gluino masses up to $650 \GeV$ are excluded for massless LSPs,
but for LSP masses above $280 \GeV$ no exclusion can be made.
In this model, LSP masses below
$200 \GeV$ are excluded for gluino masses below $600 \GeV$ and
$x > 1/2$.
The best exclusion limits are
obtained for $x = 3/4$, which gives rise to higher $\pt$ leptons
than the $x = 1/4$ case. In the squark model, no exclusion in the
$x = 1/4$ and $x = 1/2$ planes can be made.
These results are the first simplified model results in the
one-lepton channel, and complement earlier simplified model
results for the zero-lepton channel~\cite{cmsjets,cmssqgl}.

\begin{figure}
\center
\includegraphics[width=0.47\textwidth]{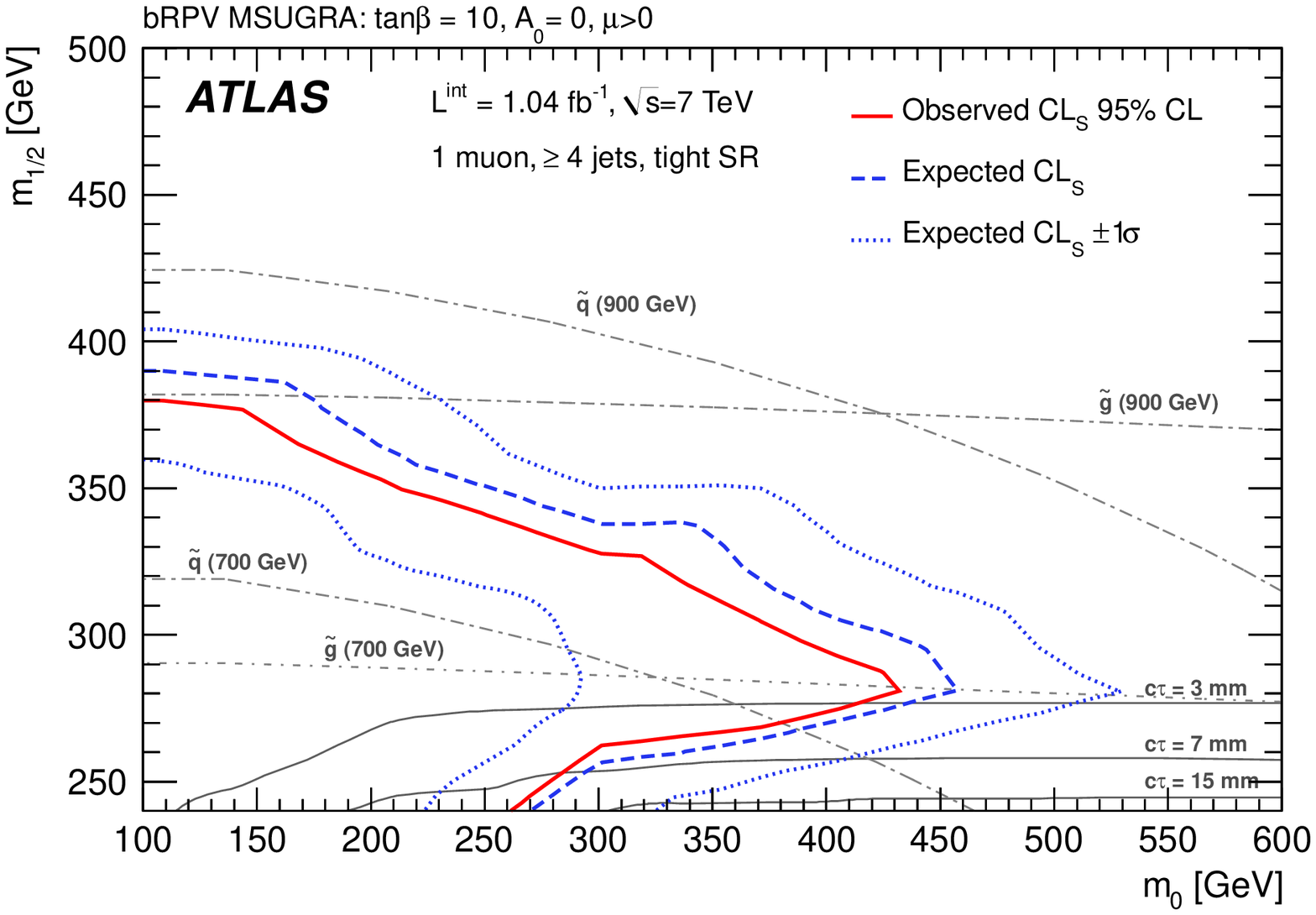}
\caption{Observed and expected 95\% CL exclusion limits, as well as the $\pm 1 \sigma$
variation on the expected limit, for the bilinear R-parity violation
model in MSUGRA parameter space using the 4JT selection in the muon channel. 
The region with LSP lifetimes $c\tau >$ 15~mm is not shown. 
}
\label{fig:brpvresult}
\end{figure}

For the bilinear R-parity violating model,
among the four signal regions considered, the tight selection 
criteria provide wider reach than the loose ones. The most stringent 
exclusion limits are set by the 4JT signal region as shown
in Figure~\ref{fig:brpvresult}.
The model is not tested for regions of parameter space where $c \tau$
of the LSP exceeds about 15 mm, which is approximately the case for
$m_{1/2} < 240 \GeV$. Within the context of this model, and for
equal squark and gluino masses, masses below $760 \GeV$ are excluded.

%
%
\section{Summary and conclusion}
\label{sec:summary}

In this paper, an update of the search for supersymmetry is presented,
in final states containing one isolated electron or muon, jets, and missing 
transverse momentum.
Good agreement is seen between the observed number of events in the
signal regions and the standard model expectation, and limits are
set on contributions of new physics to the signal regions. These
limits significantly improve on the results from 2010 data and are applied
to a wider range of SUSY models.
Model-independent limits on the cross-section of new physics contributions
to the signal regions are set, varying between 9 fb and 50 fb depending on
the channel and the signal region. In the MSUGRA/CMSSM model and
for equal squark and gluino masses, gluino masses below $820 \GeV$
are excluded. Limits are set on simplified models for gluino production and
decay and squark production and decay via an intermediate chargino.
For the gluino model and for the decay ratio $x > 1/2$, 
LSP masses below $200 \GeV$ are
excluded for gluino masses below $600 \GeV$.
For the first time at the LHC, limits are set on supersymmetric models
with bilinear R-parity violation.

%
\section{Acknowledgements}

We thank CERN for the very successful operation of the LHC, as well as the
support staff from our institutions without whom ATLAS could not be
operated efficiently.

We acknowledge the support of ANPCyT, Argentina; YerPhI, Armenia; ARC,
Australia; BMWF, Austria; ANAS, Azerbaijan; SSTC, Belarus; CNPq and FAPESP,
Brazil; NSERC, NRC and CFI, Canada; CERN; CONICYT, Chile; CAS, MOST and
NSFC, China; COLCIENCIAS, Colombia; MSMT CR, MPO CR and VSC CR, Czech
Republic; DNRF, DNSRC and Lundbeck Foundation, Denmark; ARTEMIS, European
Union; IN2P3-CNRS, CEA-DSM/IRFU, France; GNAS, Georgia; BMBF, DFG, HGF, MPG
and AvH Foundation, Germany; GSRT, Greece; ISF, MINERVA, GIF, DIP and
Benoziyo Center, Israel; INFN, Italy; MEXT and JSPS, Japan; CNRST, Morocco;
FOM and NWO, Netherlands; RCN, Norway; MNiSW, Poland; GRICES and FCT,
Portugal; MERYS (MECTS), Romania; MES of Russia and ROSATOM, Russian
Federation; JINR; MSTD, Serbia; MSSR, Slovakia; ARRS and MVZT, Slovenia;
DST/NRF, South Africa; MICINN, Spain; SRC and Wallenberg Foundation,
Sweden; SER, SNSF and Cantons of Bern and Geneva, Switzerland; NSC, Taiwan;
TAEK, Turkey; STFC, the Royal Society and Leverhulme Trust, United Kingdom;
DOE and NSF, United States of America.

The crucial computing support from all WLCG partners is acknowledged
gratefully, in particular from CERN and the ATLAS Tier-1 facilities at
TRIUMF (Canada), NDGF (Denmark, Norway, Sweden), CC-IN2P3 (France),
KIT/GridKA (Germany), INFN-CNAF (Italy), NL-T1 (Netherlands), PIC (Spain),
ASGC (Taiwan), RAL (UK) and BNL (USA) and in the Tier-2 facilities
worldwide.


\bibliographystyle{atlasnote}
\bibliography{publication_1lepton_eps2011}

\begin{thebibliography}{10}
\bibitem{atlascoordinates}
{ ATLAS uses a right-handed coordinate system with its origin at the nominal
  interaction point (IP) in the centre of the detector and the $z$-axis
  coinciding with the axis of the beam pipe. The $x$-axis points from the IP to
  the centre of the LHC ring, and the $y$ axis points upward. Cylindrical
  coordinates $(r,\phi)$ are used in the transverse plane, $\phi$ being the
  azimuthal angle around the beam pipe. The pseudorapidity is defined in terms
  of the polar angle $\theta$ as $\eta=-\ln\tan(\theta/2)$.}
\end{thebibliography}
\onecolumngrid

\clearpage
\onecolumngrid
\begin{flushleft}
{\Large The ATLAS Collaboration}

\bigskip

G.~Aad$^{\rm 48}$,
B.~Abbott$^{\rm 111}$,
J.~Abdallah$^{\rm 11}$,
A.A.~Abdelalim$^{\rm 49}$,
A.~Abdesselam$^{\rm 118}$,
O.~Abdinov$^{\rm 10}$,
B.~Abi$^{\rm 112}$,
M.~Abolins$^{\rm 88}$,
H.~Abramowicz$^{\rm 153}$,
H.~Abreu$^{\rm 115}$
% ...
\bigskip

% ...
$^{10}$ Institute of Physics, Azerbaijan Academy of Sciences, Baku, Azerbaijan\\
$^{11}$ Institut de F\'isica d'Altes Energies and Departament de F\'isica de la Universitat Aut\`onoma  de Barcelona and ICREA, Barcelona, Spain\\
% ...
$^{48}$ Fakult\"{a}t f\"{u}r Mathematik und Physik, Albert-Ludwigs-Universit\"{a}t, Freiburg i.Br., Germany\\
$^{49}$ Section de Physique, Universit\'e de Gen\`eve, Geneva, Switzerland\\
% ...
$^{88}$ Department of Physics and Astronomy, Michigan State University, East Lansing MI, United States of America\\
% ...
$^{111}$ Homer L. Dodge Department of Physics and Astronomy, University of Oklahoma, Norman OK, United States of America\\
$^{112}$ Department of Physics, Oklahoma State University, Stillwater OK, United States of America\\
% ...
$^{115}$ LAL, Univ. Paris-Sud and CNRS/IN2P3, Orsay, France\\
% ...
$^{118}$ Department of Physics, Oxford University, Oxford, United Kingdom\\
% ...
$^{153}$ Raymond and Beverly Sackler School of Physics and Astronomy, Tel Aviv University, Tel Aviv, Israel\\
% ...
\end{flushleft}


\end{document}